\documentstyle[epsfig,12pt,twoside]{article}
\begin{document}
\pagestyle{empty}
\begin{flushright}
{\small LHCb 2000-002, PHYS \\ 
(IPHE 2000-006)\\
February 11 2000}
\end{flushright}
\vspace{3cm}

\begin{center}
{\LARGE PHYSICS OF CP VIOLATION \\[3mm] AND RARE DECAYS}
\\[1cm]
{Lectures note for Cargese 99 Summer School \\
Particle Physics: Ideas and Recent Development \\
NATO Advanced Institute, Corsica, July 26-August 7, 1999}
\\[3cm]
T. NAKADA\\[5mm]
CERN 
           EP-Division\\ CH-1211 Geneva 23, Switzerland\\[1mm]
{\rm and} \\[1mm]
Institute of High
Energy Physics, University of Lausanne\\CH-1015 Dorigny,
Switzerland\\[5mm] {\small (On leave from  PSI, CH-5232 Villigen-PSI,
Switzerland)}
\end{center}
\cleardoublepage
\pagestyle{plain}
\setcounter{page}{1}
\def\ko{ {\rm K}^ 0 }
\def\kob{ \overline{\rm K}{}^0 }
\def\kl{ {\rm K_{L}} }
\def\ks{ {\rm K_{S}} }
\def\gs{ {\it \Gamma}_{\rm S} }
\def\gl{ {\it \Gamma}_{\rm L} }
\def\kos{ | {\rm K^ 0} \rangle }
\def\kobs{ | {\rm \overline{K}{}^0} \rangle }
\def\be{ \begin{displaymath} }
\def\ee{ \end{displaymath} }
\def\ben{ \begin{equation} }
\def\een{ \end{equation} }
\def\bea{ \begin{eqnarray} }
\def\eea{ \end{eqnarray} }

\section{Introduction}
Symmetries are one of the most fundamental concepts for understanding the
laws of nature leading to conserving quantities. Unexpected violations of
symmetries indicate some dynamical mechanism beyond the current
understanding of physics. 

Parity violation was discovered in 1957~\cite{bib-wu} in nuclear
$\beta$ decays and pion and muon decays~\cite{bib-gar}. In the
charged current interaction of the standard electroweak
theory, parity and charge conjugation symmetries are maximally
violated due to the
$V-A$ structure~\cite{bib-v-a}. All the experimental results up to
now are in full agreement with the theory. 

A surprising discovery of the CP violating
$\kl \rightarrow \pi^+ \pi^-$ decays~\cite{bib-chris} was made in
1964. The neutral kaon system still remains to be the only place CP
violation has been seen. The Standard Model with three Fermion families
can accommodate all the observed CP violation phenomena through the
complex quark mixing matrix, Cabibbo-Kobayashi-Maskawa (CKM) matrix~\cite{bib-ckm}. 
However, no real precision
test has been made due to the large uncertainties in evaluating
the effect of hadronic interactions.  

Interest in CP violation is not limited to elementary particle
physics. It is one of the three necessary ingredients to generate
observed excess of matter over antimatter in the
universe~\cite{nakada-sac}. The amount of CP violation which can be
generated by the Standard Model appears to be insufficient for
explaining the observed matter-antimatter asymmetry in the
universe~\cite{nakada-gab}, giving a strong motivation to search
for new physics.

For CP violation in some B meson decay channels, the Standard Model can
make precise predictions with little influence from the strong
interactions. Those channels can be used to test the
predictions quantitatively to look for a sign of new physics. Also in the
B meson system, CP violation is expected in many decay modes. The
pattern of CP violation allows us to make a systematic
qualitative comparison with the Standard Model predictions. Therefore, it
is now widely accepted that the B-meson system provides in future an ideal
place for testing the Standard Model for CP violation~\cite{sanda-carter}.

In this article, we first derive the formalism~\cite{nakada-nakada} describing the particle
antiparticle system, with and without CP violation. Three different
mechanisms which can generate CP violation are clearly classified,
together with experimental observables which identify contributions from
the different mechanisms. Then, CP violation in the neutral kaon system is
analysed in this formalism. After a brief discussion on the Standard
Model description for CP violation in the neutral kaon system, we proceed
to the neutral B meson system. Following the discussion on some Standard
Model predictions, some thoughts are made how the situation could change
if there exists new physics contributing in the B meson system. 

\section{Description of Particle Antiparticle System}
\def\po{ {\rm P}^ 0 }
\def\pob{ \overline{\rm P}{}^0 }
\def\pmi{ {\rm P_{-}} }
\def\ppl{ {\rm P_{+}} }
\def\gp{ {\it \Gamma}_{+} }
\def\gm{ {\it \Gamma}_{-} }
\def\pos{ | {\rm P^ 0} \rangle }
\def\pobs{ | {\rm \overline{P}{}^0} \rangle }
\def\itagamma{{\it \Gamma}}
\def\italambda{{\it \Lambda}}
\subsection{Basic Formalism}
Let $\pos$ and $\pobs$ be the states of a neutral
pseudoscalar particle $\po$-meson and its antiparticle $\pob$-meson at
rest, respectively. They have definite flavour quantum numbers with
opposite signs: $F=+1$ for
$\po$ and $F=-1$ for $\pob$. Both states are eigenstates of the strong and
electromagnetic interaction Hamiltonian, i.e.  
\be
\left( H_{\rm st}+H_{\rm em} \right) \pos = m_0 \pos~~{\rm and}~~
\left( H_{\rm st}+H_{\rm em} \right) \pobs = \overline{m}{}_0 \pobs
\ee
where $m_0$ and $\overline{m}{}_0$ are the rest masses of $\po$ and
$\pob$, respectively. 
The $\po$ and $\pob$ states are related through CP transformations. For
stationary states, the T transformation does not alter them, with the
exception of an arbitrary phase. While CP is a unitary operation, T is an
antiunitary operation. 

In summary, we obtain
\ben
\begin{array}{c} 
C\!P \, \pos = e^{i \, \theta _{\rm CP} } \pobs~~{\rm and}~~
C\!P \, \pobs = e^{- \,i \, \theta_{\rm CP} } \pos~~ \\
T \, \pos = e^{ i \, \theta_{\rm T} } \pos~~{\rm and}~~
T \, \pobs = e^{ i \, \overline {\theta}{}_{\rm T } } \pobs
\end{array} \label{cp-phase}\een
where the $\theta$'s are arbitrary phases, and by assuming $C\!PT\, \pos
= TC\!P \, \pos$ it follows that 
\be
2 \, \theta_{\rm CP} =\overline{\theta}{}_{\rm T}\, - \, \theta_{\rm
T}~.
\ee

Since T is antiunitary, it follows that 
\be
T\, c = c^*\, T
\ee
where $c$ is any complex number. If we define
\be
T |\alpha\rangle =|\tilde{\alpha}\rangle,~ T |\beta \rangle =|
\tilde{\beta}\rangle 
\ee
antiunitary operation has to give
\be
\langle \alpha | \beta \rangle = \left[\langle \tilde{\alpha} |
\tilde{\beta}
\rangle \right]^*~.
\ee
On the other hand, 
\be
\langle \alpha | \beta \rangle=\langle \alpha | \left(T^{-1}T|\beta
\rangle\right)=\langle \alpha | \left(T^{-1}|\tilde{\beta}
\rangle\right),
\ee
hence
\be
\langle \alpha | \left(T^{-1}|\tilde{\beta}
\rangle\right)=\left[\langle \tilde{\alpha} |
\tilde{\beta}
\rangle \right]^*~.
\ee
We can then conclude
\be
\langle \alpha | \left(T^{-1}|\tilde{\beta}
\rangle\right)=\left[ \left(\langle \alpha | T^{-1}\right)|\tilde{\beta}
\rangle \right]^*
\ee
i.e. when the T operator changes the direction of the operation, it must
be complex conjugated.

If strong and electromagnetic interactions are
invariant under the CPT transformation, which is assumed throughout this
paper, it follows that $m_0=\overline{m}{}_0$.

Now we switch on an interaction, $V$, and
the P can decay into final states f with different flavours ($|\Delta
F|= 1$ process) and $\po$ and $\pob$ can oscillate to each other
($|\Delta F| = 2$ process). Thus, a general
state $| \psi (t) \rangle$ which is a solution of the Schr\"odinger
equation
\ben
i \, \frac{\partial}{\partial t} |\psi (t)\rangle
 = \left( H_{\rm st} + H_{\rm em} + V \right) | \psi (t) \rangle
\label{schrodinger} \een
can be written as
\be
| \psi(t) \rangle = a(t) \pos + b(t) \pobs + \sum_{\rm f} c_{\rm f}
(t)| {\rm f} \rangle 
\ee
where the sum is taken over all the possible final states f and 
$a(t)$,
$b(t)$ and
$c_{\rm f}(t)$ are time dependent functions; $|a(t)|^2$, $|b(t)|^2$ and
$|c_{\rm f}(t)|^2$ give the fractions of $\po$, $\pob$ and f at time $t$
respectively. Since the weak interaction is much weaker than strong and
electromagnetic interactions, perturbation theory can be applied in order
to solve equation \ref{schrodinger}. Also with the help of the
Wigner-Weisskopf method, which neglects the weak interactions between the
final states~\cite{wiger-weis}, and we obtain
\ben
i \, \frac{ \partial }{\partial t} 
\left( \begin{array}{c}a(t)\\b(t)\end{array}\right) =
\mbox{\boldmath$\Lambda$} \left(
\begin{array}{c}a(t)\\b(t)\end{array}\right)  = \left(
\mbox{\boldmath$M$} -
\, i \, \frac{\, \mbox{\boldmath$\Gamma$}\,}{2} \right) \left (
\begin{array}{c}a(t)\\b(t)\end{array}\right)
\label{basic}
\een
where the $2\times 2$ matrices {\boldmath$M$} and
{\boldmath$\Gamma$} are often referred to as the mass and
decay matrices. 

The elements of the mass matrix are given as
\ben
M_{ij}= m_{0} \, \delta_{ij} + \langle i |V| j \rangle 
+\sum_{\rm f}{\cal P}\left (\frac{\langle i|V|{\rm f} \rangle \langle
{\rm f} |V| j\rangle }{m_0-E_{\rm f}}\right ) \label{m12}
\een
where ${\cal P}$ stands for the principal part and the index $i=1$(2)
denotes $\po$($\pob$). Note that the sum is taken over {\it all possible
intermediate states} common to $\po$ and $\pob$ for $i\neq j$. 

The elements of the decay matrix are given by 
\ben  {\it \Gamma}_{ij}=2\,
\pi\sum_{\rm f}\langle i|V|{\rm f} \rangle \langle {\rm
f}|V|j\rangle \delta(m_0-E_{\rm f}) 
\label{g-element}\een
The sum is taken over only {\it real final states} common to $\po$ and
$\pob$ for $i\neq j$.  

If the Hamiltonians are not Hermitian, transition probabilities are
not conserved in decays or oscillations, i.e. the number of initial
states is not identical to the number of final states. This is
also referred to as the break down of unitarity. We assume from now
on that all the Hamiltonians are Hermitian, i.e.
\be
|a(t)|^2 + |b(t)|^2 + \sum_{\rm f} |c_{\rm f}|^2 = 1,
\ee
and also 
\be
M_{ij}=M_{ji}^*,~{\it \Gamma}_{ij}={\it \Gamma}_{ji}^*~.
\ee
Clearly $|a(t)|^2 + |b(t)|^2$ decreases as a function of time, hence
{\boldmath$\Lambda$} is not Hermitian.

Since the CP operator changes a particle state into an antiparticle
state, the following relation can be obtained if $V$ is invariant under
the CP transformation, i.e. $(CP)^{-1}\, \, V\, \, CP=V$:
\be
{\rm CP}:~~\left| {\it \Lambda}_{12}
\right| = \left| {\it \Lambda}_{21} \right|,~{\it \Lambda}_{11}={\it
\Lambda}_{22}~.
\ee
Since the T operator induces complex conjugation, which is equivalent
to interchanging a bra-state and a ket-state, the following relation can
be obtained if $V$ is invariant under the T transformation: 
\be
{\rm T}:~~\left| {\it \Lambda}_{12}
\right| = \left| {\it \Lambda}_{21} \right|~.
\ee
By combining the two, we obtain for the CPT invariant case:
\be
{\rm CPT}:~~{\it \Lambda}_{11}={\it
\Lambda}_{22}~.
\ee
For a rigorous proof,
equations \ref{cp-phase}, \ref{m12} and
\ref{g-element}  are used. 

It follows that 
\ben
\begin{array}{l}
\bullet {\rm if}~{\it \Lambda}_{11}\neq{\it \Lambda}_{22},~{\rm
i.e.}~M_{11}
\neq M_{22}~{\rm or}~ {\it
\Gamma}_{11}\neq {\it
\Gamma}_{22}:\\
~~{\rm{ \bf CPT~and~CP}~are~violated}
\vspace{2 mm}
\\
\bullet {\rm if}~|{\it \Lambda}_{12}|\neq|{\it \Lambda}_{21}|,~{\rm
i.e.}~\sin \left(\varphi_{\it \Gamma} -
\varphi_{M}\right) \neq 0:\\ ~~
{\rm {\bf T}~{\bf and~CP}~are~violated}~.
\label{CPCPT}
\end{array}
\een
where $\varphi_{M}=\arg\left( M_{12} \right)~~{\rm and}~~ \varphi_{\it
\Gamma}=\arg\left( {\it \Gamma}_{12}\right )$. Note that CP violation in
the mass and decay matrices cannot be separated from CPT violation or T
violation. 

While there is no fundamental reason to respect CP and T
symmetries, it can be shown based on only few
basic assumptions that no self consistent quantum field theory can be
constructed that does not conserve CPT symmetry~\cite{cite-cpt}.
Therefore, we restrict our further discussion to the case where CPT
symmetry is conserved: i.e.
\be
M_{11}=M_{22}\equiv M,~{\it \Gamma}_{11}={\it \Gamma}_{22}\equiv {\it
\Gamma}
\ee
thus
\be
{\it \Lambda}_{11}={\it \Lambda}_{22}\equiv {\it \Lambda}~.
\ee

Differential equation \ref{basic} can be reduced to
\ben
\frac{d^2\,a(t)}{dt^2}+2\, i\, {\it \Lambda}\, \frac{d\, a(t)}{dt} +
\left( {\it
\Lambda}_{12} {\it \Lambda}_{21} - {\it \Lambda}^2 \right) a(t)=0
\label{deffeq}\een
for $a(t)$, and a general solution is given by
\be
a(t)=C_+e^{-i \lambda_+}+C_-e^{-i \lambda_-}
\ee
where $C_{\pm}$ are arbitrary constants which can only be defined by
the initial condition. For $b(t)$, we obtain 
\be
b(t)=\frac{1}{{\it \Lambda}_{12}} \left[ i\, \frac{d\, a(t)}{dt} - {\it
\Lambda}\, a(t) \right]
\ee
which can be used once $a(t)$ becomes known.

Insertion of 
$a(t)$ into equation
\ref{deffeq} leads to
\be
\lambda_{\pm}^2 - 2 \italambda \lambda_\pm +\left( \italambda_{12}
\italambda_{21} - \italambda^2 \right) = 0 
\ee
from which the eigen-frequencies are obtained as
\be
\lambda_{\pm}={\it \Lambda}\pm \sqrt{{\it \Lambda}_{12} {\it
\Lambda}_{21}}\equiv m_\pm - \frac{i}{2}{\it \Gamma}_\pm
\ee
by solving
where 
\ben
m_\pm = \Re{\lambda_\pm} = M \,
\pm \, \Re  \left( \, {\it \Lambda}_{12}\,{\it
\Lambda}_{21}\,\, \right)^{1/2} \label{mass-1}
\een
and
\ben 
{\it \Gamma}_\pm = - 2\, \Im
{\lambda_\pm} = {\it
\Gamma} \, \mp \,2\,
\Im
\left(
\, {\it \Lambda}_{12}\,{\it
\Lambda}_{21}\,\, \right)^{1/2}~. \label{gamma-1}
\een 

For an initially pure $\po$ state, we have $a(t)=1$ and $b(t)=0$ at
$t=0$, i.e. $C_+=C_-=1/2$, and the solution is given by
\bea 
|{\rm P}{}^0(t)\rangle &=& a(t) \pos + b(t) \pobs \nonumber \\
&=& f_+(t) \pos  + \zeta f_-(t)\, \pobs
\label{k0state}  \label{eq-po1}\\  
&=&\frac{\,\sqrt{\, 1+|\zeta |^2 \,}\,}{2} \left(\, |\ppl
\rangle\, e^{-\, i\,
\lambda_{+}\, t } + |\pmi \rangle\, e^{-\, i\, \lambda_{-}\, t }\,
\right)
\label{eq-po2} \eea 
where 
\be
f_{\pm}(t)=\frac{1}{\,2\,}\left(\, e^{-\, i\, \lambda_{+}\, t }
\pm
 e^{-\, i\, \lambda_{-}\, t}\, \right)
\ee
and $\zeta$ is
\ben
\zeta = \sqrt{ \, \frac{\, {\it \Lambda}_{21}\,}{{\it \Lambda}_{12}}\,}~.
\label{zeta}
\een 
The two states $| \ppl \rangle $ and $| \pmi \rangle$
are the eigenstates of $\lambda_\pm$ and are given by
\ben
|{\rm P}_{\pm} \rangle = \frac{\displaystyle 1}{\displaystyle \,
\sqrt{\,1+|\zeta |^2 \,} \,}
\left(\pos \, \pm \, \zeta \,\pobs \, \right)~.
\label{klks}\een
For an initially pure $\pob$ state, we have 
\bea 
|\overline{\rm P}{}^0(t)\rangle &=& \frac{1}{\, \zeta \,}f_-(t) \,
\pos  +  f_+(t) \, \pobs \label{eq-pob1}\\ 
&=&\frac{\,\sqrt{\, 1+|\zeta |^2 \,}}{2\,\zeta } \left(\,|\ppl
\rangle\, e^{-\, i\,
\lambda_{+}\, t } -|\pmi \rangle\, e^{-\, i\, \lambda_{-}\, t }\,
\right)~.
\label{eq-pob2} \eea 

While $\rm P^\pm$ have definite masses and decay widths (as seen from
equations \ref{eq-po2} and \ref{eq-pob2}), $\po$ and $\pob$ do not and
they oscillate to each other (see equations \ref{eq-po1} and
\ref{eq-pob1}).

\subsection{CP Conserving Case}
If $V$ remains invariant under the CP transformation, from equations
\ref{cp-phase}, \ref{m12} and \ref{g-element} it follows that
\be
M_{12}=M_{21}e^{-i\, 2\,\theta_{\rm CP}}
=M_{12}^*e^{-i\, 2\,\theta_{\rm CP}}
\ee
thus
\be
\arg M_{12}= -\theta_{\rm CP} + n\pi,
\ee
and
\be
{\it \Gamma}_{12}={\it \Gamma}_{21}e^{-i\, 2\,\theta_{\rm CP}}
={\it \Gamma}_{12}^*e^{-i\, 2\,\theta_{\rm CP}}
\ee
thus
\be
\arg \itagamma_{12}= -\theta_{\rm CP} + n'\pi,
\ee
where $n$ and $n'$ are arbitrary integer numbers.

For $\zeta$, we have 
\be
\zeta = \sqrt{\frac{{\it \Lambda}_{21}}{{\it \Lambda}_{12}}} = e^{i
\, (\theta_{\rm CP} + n'' \pi)}
\ee
where $n''$ is an arbitrary integer number. The two mass eigenstates
$|{\rm P}_{+} \rangle
$ and $|{\rm P}_{-} \rangle $ become CP eigenstates
\be
CP |{\rm P}_\pm \rangle = \pm\, (-1)^{n''} |{\rm P}_\pm \rangle~.
\ee
The mass and decay width eigenvalues, equations
\ref{mass-1} and \ref{gamma-1}, become
\be
m_{\pm} = M \pm (-1)^{n+n''}|M_{12}|
\ee
and
\be
{\it \Gamma}_{\pm }={\it \Gamma} \pm (-1)^{n'+n''}|{\it \Gamma}_{12}|
\ee

\begin{figure}
\begin{center}
\epsfig{file=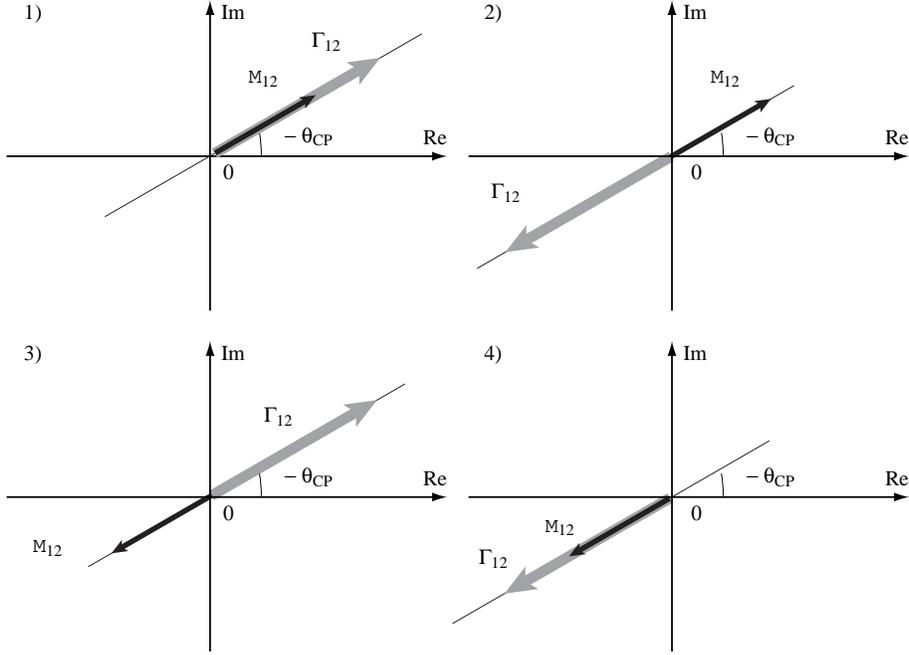,width=120mm}
\end{center}
\caption{Relative phase relations for $M_{12}$, $\itagamma_{12}$, and CP
transformation phase $\theta_{\rm CP}$ when CP is conserved: 1)~ $CP=+1$
state is heavier and decays faster, 2)~$CP=+1$ state is heavier and
decays slower, 3)~$CP=+1$ state is lighter and
decays faster, 4)~ $CP=+1$ state is lighter and
decays slower. }
\label{fig-phase}
\end{figure}

By examining various combinations of $n$, $n'$ and $n''$, we can
show that the following four possibilities exist:
\begin{enumerate}
\item $n$=even, $n'$=even: $CP=+1$ state is heavier and
decays faster,
\item $n$=even, $n'$=odd: $CP=+1$ state is heavier and
decays slower,
\item $n$=odd, $n'$=even: $CP=+1$ state is lighter and
decays faster,
\item $n$=odd, $n'$=odd: $CP=+1$ state is lighter and
decays slower.
\end{enumerate}
Figure~\ref{fig-phase} illustrates the phase relations in a pictorial
way.  The choice of $n''$ does not alter the conclusion and $n''=0$ can be
adopted without any loss of generality. In this case, $|{\rm P}_{+}
\rangle $ is the $CP=+1$ state.

\subsection{CP Violating Case}
Let us consider the time dependent decay rate for the initial $\po$
decaying into a CP eigenstate f, given by $| \langle f |V| \po(t)
\rangle |^2$, and that for the initial $\pob$ decaying into f, given by $|
\langle f |V| \pob(t)
\rangle |^2$:
\bea
R_{\rm f}(t) &\propto&
|f_+(t)|^2 + \left|\zeta  \frac{\overline{A}{}_{\rm f}}{A_{\rm f} }
\right|^2 |f_-(t)|^2 +2\Re \left[\zeta\frac{\overline{A}{}_{\rm
f}}{A_{\rm f}}
\, f_+^*(t)f_-(t) \right]
\label{rate1} \\
\overline{R}{}_{\rm f}(t) &\propto&
\left| \frac{\overline{A}{}_{\rm f}}{A_{\rm f}}\right|^2 |f_+(t)|^2 +
\left| \frac{1}{\zeta}\right|^2 |f_-(t)|^2 +\frac{2}{|\zeta|^2}\Re
\left[\zeta^* \frac{\overline{A}{}_{\rm f}^* }{A_{\rm f}^*} 
f_+^*(t)f_-(t) \right]
\label{rate2}\eea
where the instantaneous decay amplitudes are denoted by
$A_{\rm f}\equiv \langle {\rm f}| V \pos$ etc. and equations \ref{eq-po1}
and \ref{eq-pob1} are used. 

Since $R_f(t)$ and $\overline{R}{}_{\rm f}(t)$ describe the CP
conjugated processes to each other, any difference between the two is an
clear proof of CP violation. As seen from the first terms of
equations~\ref{rate1} and \ref{rate2}, CP violation is generated if
$|A_{\rm f}|\neq |\overline{A}{}_{\rm f}|$. This is called
{\bf CP violation in the decay amplitudes}.

From the second terms of $R_{\rm f}(t)$ and $\overline{R}{}_{\rm
f }(t)$, it can be seen that CP violation is generated if
$|\zeta| \neq 1$ even if there is no CP violation in the decay
amplitudes. From equations
\ref{eq-po2} and
\ref{eq-pob2}, it is clear that the oscillation rate for $\po \rightarrow
\pob$ is different from that for $\pob \rightarrow \po$ if $|\zeta|\neq1$,
thus this is called {\bf CP violation in the oscillation}.

The third term can be expanded into 
\be
2\Re \left( \zeta \frac{ \overline{A}{}_{\rm f}}{A_{\rm f}} \right)
\Re \left[ f_+^*(t)f_-(t) \right] - 
2\Im \left(  \zeta \frac{ \overline{A}{}_{\rm f}}{A_{\rm f}} \right)
\Im \left[ f_+^*(t)f_-(t) \right] 
\ee
for $R_{\rm f}(t)$ and 
\be
\frac{2}{|\zeta|^2} \Re \left( \zeta \frac{
\overline{A}{}_{\rm f}}{A_{\rm f}}
\right)
\Re \left[ f_+^*(t)f_-(t) \right] + 
\frac{2}{|\zeta|^2}\Im \left( \zeta \frac{
\overline{A}{}_{\rm f}}{A_{\rm f}} \right)
\Im \left[ f_+^*(t)f_-(t) \right] 
\ee
for $\overline{R}{}_{\rm f}(t)$. If CP violation in $\po$-$\pob$
oscillation is absent, the first terms are identical. Even in that case,
if
\be
\Im \left( \zeta \frac{
\overline{A}{}_{\rm f}}{A_{\rm f}} \right)\neq 0
\ee 
CP violation is still present. Since the process involves the decays of
$\po$ ($\pob$) from the initial
$\po$ ($\pob$) and decays of the $\pob$ ($\po$) oscillated from the
initial $\po$ ($\pob$) into a common final state, it is referred as {\bf
CP violation due to the interplay between the decays and oscillations}.

If \label{page-approx}CP violation in $\po$-$\pob$ oscillation is small,
i.e.
$\left( |\zeta|-1 \right)^2<<1$, we can derive  
\be
|\sin (\varphi_{\it \Gamma}-\varphi_{M})|<<1
\ee
from equation \ref{zeta}, where $\varphi_{\it \Gamma}=\arg {\it
\Gamma}_{12}$ and $\varphi_{M}=\arg M_{12}$ as already defined. By
introducing
$|\Delta_{\itagamma/M}| <<1 $ as
\ben
\varphi_{\it \Gamma}-\varphi_{M}=n\, \pi-\Delta_{\itagamma/M}
\label{ndef}\een
where $n$ is an integer number, the following two approximations are
possible:
\\
\noindent a) $\varphi_\itagamma = \arg \itagamma_{12}$ base
\bea
\zeta &\approx& \left\{ 1 - \frac{2 |M_{12}||\itagamma_{12}|
\Delta_{\itagamma/M}}{4|M_{12}|^2+|\itagamma_{12}|^2} \left[ (-1)^{n+1} +i
\frac{2 |M_{12}|}{|\itagamma_{12}|} \right] \right\}
e^{- i\, \varphi_\itagamma} \nonumber \\[2mm]
m_{\pm}&=&M \pm (-1)^n|M_{12}| \nonumber \\
\itagamma_{\pm} &=& \itagamma \pm |\itagamma_{12}|
\nonumber 
\eea
b) $\varphi_M  = \arg M_{12}$ base
\bea
\zeta &\approx& \left\{ 1 + \frac{2 |M_{12}||\itagamma_{12}|
\Delta_{\itagamma/M}}{4|M_{12}|^2+|\itagamma_{12}|^2} \left[ (-1)^n +i
\frac{|\itagamma_{12}|}{2 |M_{12}|} \right] \right\}
e^{- i\, \varphi_M}
\label{mbase} \\[2mm]
m_{\pm}&=&M \pm |M_{12}| \nonumber \\
\itagamma_{\pm} &=& \itagamma \pm (-1)^n|\itagamma_{12}|~.
\nonumber 
\eea


\section{Neutral Kaon System}
\subsection{Adaptation of Formalism}
Now we adapt the above developed formalism to the neutral kaon system. As
described later, observed CP violation in the $\ko$-$\kob$ oscillation
is very small.  The two mass eigenstates are called
$\ks$ and
$\kl$ with corresponding masses and decay widths referred to as $m_{\rm
S}$,
$m_{\rm L}$, ${\it
\Gamma}_{\rm S}$ and ${\it \Gamma}_{\rm L}$ respectively and they are
known to be
$m_{\rm S}<m_{\rm L}$ and ${\it \Gamma}_{\rm S}>{\it \Gamma}_{\rm L}$.
Therefore, $M_{12}$ and $\itagamma_{12}$ is {\bf almost antiparallel to
each other}, thus $n=1$ in equation \ref{ndef}.

Since the kaon decay properties are experimentally well measured,
enough information is available to calculate $\itagamma_{12}$ from the
data, as described in Section~\ref{sec-gamma}. We therefore adopt the
$\varphi_{\itagamma}$ base given in the previous section. 

It follows that
\bea
\zeta = (1-2 \epsilon) e^{-i \varphi_{\it \Gamma}}
\label{zetak}\eea
where the small parameter $\epsilon$ is given by
\be
 \epsilon= \frac{\, |M_{12}| |{\it \Gamma}_{12}|\sin \left(\varphi_{\it
\Gamma} - \varphi_{M} \right)\,}{ 4|M_{12}|^2 + |{\it
\Gamma}_{12}|^2 } \left( 1+ i \frac{\, 2 |M_{12}| \,}{|{\it
\Gamma}_{12}|} \right)~.
\ee
and
\bea
|\ks \rangle &=& \frac{1}{\, \sqrt{1+|\epsilon|^2}\,} \left [\kos + (1-2
\epsilon)  e^{-i \varphi_{\it \Gamma}} \kobs \right]
\label{ksdef} \\
|\kl \rangle &=& \frac{1}{\, \sqrt{1+|\epsilon|^2}\,} \left [\kos - (1-2
\epsilon)  e^{-i \varphi_{\it \Gamma}} \kobs \right]~.
\label{kldef} \eea
From the measured lifetimes~\cite{nakada-pdg},
\be
\tau_{\rm s} \equiv \frac{1}{\, \gs \,} = (0.8934 \pm 0.0008) \times
10^{-10}~{\rm s}
\ee
and
\be
\tau_{\rm L} \equiv \frac{1}{\, \gl \,} = (5.17 \pm 0.04) \times
10^{-8}~{\rm s}
\ee
i.e.
\be
\Delta {\it \Gamma}={\it \Gamma}_{\rm S} - {\it \Gamma}_{\rm L} =
(1.1174 \pm 0.0010) \times 10^{10}~{\rm s}^{-1}
\ee
and
the mass difference, 
\be
\Delta m \equiv m_{\rm L}- m_{\rm S} = (0.5301 \pm 0.0014)
\times 10^{10}~\hbar {\rm s}^{-1}
\ee
we obtain,
\be
\frac{ |M_{12}| |{\it \Gamma}_{12}|}{ 4|M_{12}|^2 + |{\it
\Gamma}_{12}|^2 }=0.24966\pm0.00004
\ee
and
\be
 \frac{\, 2 |M_{12}| \,}{|{\it
\Gamma}_{12}|}=0.9488 \pm 0.0026~.
\ee

Since the lifetime of $\kl$ is much longer than that of $\ks$, it is
possible to produce a $\kl$ beam. Therefore, many kaon experiments have
been done using $\kl$ beams.

\subsection{CP Violation in Oscillations}
\begin{figure}
\begin{center}
\epsfig{file=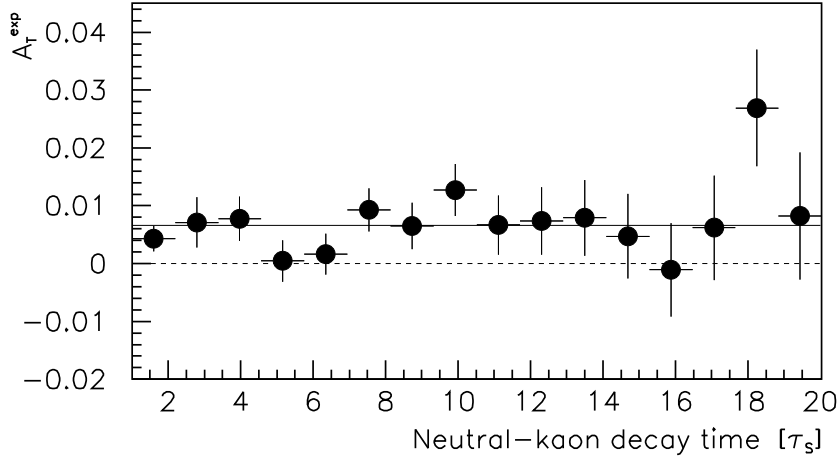,width=110mm}
\end{center}
\caption{Measured rate asymmetry between the initial $\kob$ decaying into
$\rm e^+ \pi^- \nu$ and the initial $\ko$ decaying into $\rm e^-
\pi^+\overline{\nu}$ as a function of the decay time in units of
$\tau_{\rm S}$ by the CPLEAR experiment. The solid line is
obtained by fitting a constant value.}
\label{fig-at}
\end{figure}

The CPLEAR experiment observed CP violation in the $\ko$-$\kob$
oscillation by measuring the difference in the oscillation rates
between $\kob \rightarrow \ko$ and $\ko \rightarrow \kob$. The initial
neutral kaons were produced by $\rm p
\overline{p}$ annihilations:
$\rm p \overline{p}
\rightarrow K^0 K^- \pi^+$ and $\rm \rightarrow \overline{K}{}^0 K^+
\pi^-$, where the initial flavour can be defined by the charge sign of the
accompanying kaon. Semileptonic decays were used in order to determine the
flavour at the moment of the decay. Since the $\ko$ contains an $\rm
\overline{s}$-quark (and $\kob$ an s-quark), $\ko$ ($\kob$) can decay only
into
$\rm e^+ \pi^- \nu$ ($\rm e^- \pi^+ \overline{\nu}$)
instantaneously. Therefore, the initial $\ko$ ($\kob$) can produce the
final state
$\rm e^- \pi^+ \overline{\nu}$ ($\rm e^+ \pi^- \nu$) only through the
$\ko \rightarrow \kob$ ($\kob \rightarrow \ko$) oscillation. From the two
measured time dependent decay rates,
$R_{\rm e^-}(t)$ and $\overline{R}_{\rm e^+}(t)$, an asymmetry
\be
A_{\rm T}(t) =\frac{\overline{R}_{\rm e^+}(t) - R_{\rm
e^-}(t)}{\overline{R}_{\rm e^+}(t) + R_{\rm e^-}(t)}
\ee
is constructed as shown in Figure~\ref{fig-at}. Using equations
\ref{eq-po1}, 
\ref{eq-pob1} and \ref{zetak}, it follows that
\be
A_{\rm T}(t)=\frac{1-|\zeta|^4}{\, 1+|\zeta|^4\, } = 4\, \Re \epsilon
\ee
and from the measured $A_{\rm T}(t)=(6.6\pm 1.6) \times
10^{-3}$~\cite{Angelopoulos:1998dv},
\be
|\zeta|=0.9967\pm 0.0008\neq 1
\ee 
is obtained exhibiting a clear
sign of CP violation and T violation in the
$\ko$-$\kob$ oscillation. 

The parameter $|\zeta|$ can also be measured from the semileptonic
branching fractions of $\kl$ by the lepton sign asymmetry: using equations
\ref{kldef} and \ref{zetak}, we obtain~\cite{nakada-pdg} 
\bea
\delta_{\ell} &\equiv& \frac{B(\kl \rightarrow \ell^+ \pi^- \nu) -
B(\kl \rightarrow \ell^- \pi^+ \overline{\nu}) }
{B(\kl \rightarrow \ell^+ \pi^- \nu) +
B(\kl \rightarrow \ell^- \pi^+ \overline{\nu}) } \nonumber \\
&=&\frac{1-|\zeta|^2}{1+|\zeta|^2} = 2\Re \epsilon
\nonumber \\
&=&(3.27\pm0.12)\times 10^{-3}
\nonumber \eea
where $\ell$ can be e or $\mu$ and $B$ stands for a branching fraction.

Using all the measurements, we obtain
\be
\Re \epsilon = (1.64 \pm 0.06) \times 10^{-3}
\ee
and
\be
\arg{\epsilon}=(43.50\pm0.08)^\circ .
\ee

\subsection{CP Violation due to Decays and Oscillations}
Since the two-pion final state is a CP eigenstate with $CP=+1$, $\kl$
decaying into $\pi^+ \pi^-$ is a CP violating decay. This was indeed the
first observed sign of CP violation. A commonly used CP violation
parameter
$\eta_{+-}$ is defined as
\ben
\eta_{+-}\equiv
\frac{\langle \pi^+ \pi^-
| V | \kl \rangle }{\langle \pi^+ \pi^-
| V | \ks
\rangle}=\frac{1-\zeta\frac{\displaystyle
\overline{A}{}_{+-}}{\displaystyle A_{+-}}}{\, 1+\zeta\frac{\displaystyle
\overline{A}{}_{+-}}{\displaystyle A_{+-}}\, }
\label{etappdef}\een
where equations \ref{ksdef}, \ref{kldef} are used and $A_{+-}$ and
$\overline{A}{}_{+-}$ denote the $\ko$ and $\kob \rightarrow \pi^+ \pi^-$
decay amplitudes respectively.

\begin{figure}
\begin{center}
\epsfig{file=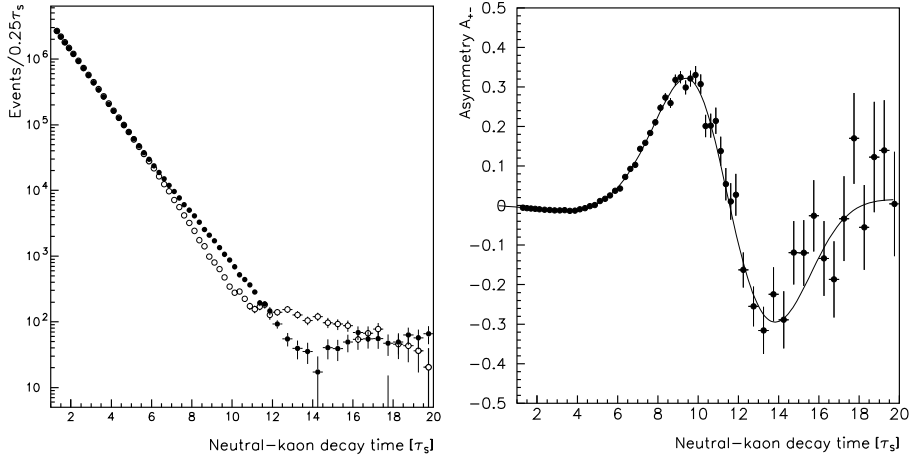,width=120mm}
\end{center}
\caption{The time dependent rate distributions for the initial $\kob$
(solid circles) and
$\ko$ (open circles) decaying into
$\pi^+ \pi^-$ as a function of the decay time in units of
$\tau_{\rm S}$ obtained by the CPLEAR experiment. The rate asymmetry is
also shown.}
\label{fig-ktime}
\end{figure}

The parameter $\eta_{+-}$ can be measured from 
the time dependent decay rates for the initial $\ko$ and $\kob$ into
$\pi^+ \pi^-$. From equations \ref{eq-po2} and \ref{eq-pob2}, the two
rates are given by
\be
R_{+-}(t) \propto \frac{1}{2}
e^{- {\it \Gamma}_{\rm S}\, t} +
\left| \eta_{+-} \right|^2 e^{- {\it \Gamma}_{\rm L}\, t}
+ 2 |\eta_{+-}| e^{- \hat{\it \Gamma}\,t} \cos (\Delta m \, t - 
\phi_{+-} )
\ee
and
\be
\overline{R}_{+-}(t) \propto \frac{\, 1+4 \Re \epsilon \,}{2}
\left[ 
e^{- {\it \Gamma}_{\rm S}\, t} +
\left| \eta_{+-} \right|^2 e^{- {\it \Gamma}_{\rm L}\, t}
- 2 |\eta_{+-}| e^{- \hat{\it \Gamma}\,t} \cos (\Delta m \, t - 
\phi_{+-} ) \right]
\ee
where $\phi_{+-}$ is the phase of $\eta_{+-}$ and $\hat{\it \Gamma}$ is
the $\ks$-$\kl$ average decay width. The second term is CP violating $\kl$
decays and the third term is due to the interference between the $\ks$
decay and CP violating
$\kl$ decay amplitudes. Figure~\ref{fig-ktime}
shows~\cite{Apostolakis:1999zw} the measured
$R_{+-}(t)$ and
$R_{+-}(t)$ together with the CP asymmetry defined as
\be
A_{+-}(t)=\frac{\overline{R}_{+-}(t)-R_{+-}(t)}{\overline{R}_{+-}(t)+R_{+-}(t)}
\ee
where the interference term is well isolated. At around $t=10 \tau_{\rm
S}$, the $\ks$ decay rate is reduced to the level of the CP violating
$\kl$ decay rate, thus the asymmetry becomes very large.

This direct comparison between the two CP conjugated processes illustrates
another straightforward demonstration of CP violation in the neutral kaon
system. From the asymmetry, the value of
$\eta_{+-}$ is measured to be~\cite{Apostolakis:1999zw}
\be
|\eta_{+-}|=(2.264\pm0.035)\times 10^{-3},~
\phi_{+-}=(43.19\pm0.60)^{\circ}
\ee
which leads to 
\be
\Im \left(  \zeta\frac{\displaystyle
\overline{A}{}_{+-}}{\displaystyle A_{+-}} \right) =-
(3.099\pm0.048)\times 10^{-3}
\ee
exhibiting that CP violation due to the interference between the decay
and oscillation is present.

\subsection{CP Violation in Decays}
The two-pion final state can be in a total isospin state of $I=0$
or $I=2$. The $I=1$ state is not allowed due to Bose statistics. Using the
isospin decomposition, we can derive the $\ko$ and $\kob$ decay
amplitudes to $\pi^+ \pi^-$ to be
\be
A_{+-}=
\sqrt{ \frac{2}{3} } \langle 2\pi(I=0)| V |\ko  \rangle +
\sqrt{ \frac{1}{3}} \langle 2\pi(I=2)| V |\ko  \rangle
\ee
and
\be
\overline{A}{}_{+-}=
\sqrt{ \frac{2}{3} } \langle 2\pi(I=0)| V |\kob  \rangle +
\sqrt{ \frac{1}{3}} \langle 2\pi(I=2)| V |\kob  \rangle~.
\ee
Using CPT symmetry and the S-matrix, the $\ko$ and $\kob$ decay amplitudes
can be related and it follows that 
\bea
A_{+-}&=&\sqrt{ \frac{2}{3} } a_0 e^{i\, \delta_0} +
\sqrt{ \frac{1}{3} } a_2 e^{i\, \delta_2} \nonumber \\
\overline{A}{}_{+-}&=&\sqrt{ \frac{2}{3} } a_0^* e^{i\,\left( \delta_0 + \theta_{\rm
CP}-\overline{\theta}{}_{\rm T}\right)} +
\sqrt{ \frac{1}{3}} a_2^* e^{i\,\left( \delta_2 + \theta_{\rm
CP}-\overline{\theta}{}_{\rm T}\right)}
\nonumber\eea
where $a_0$ and $a_2$ are the $\ko$ decay amplitudes into $2\pi(I=0)$ and
$2\pi(I=2)$ states due to the short-range weak interactions and
$\delta_0$ and $\delta_2$ are the $\pi$-$\pi$ scattering phase
shifts for the $I=0$ and $I=2$ two-pion configuration at $\sqrt{s}=m_{\rm
K}$ respectively. It is important to note that the two-pion scattering is
totally dominated by the elastic scattering at the energy scale of the kon
mass. Similarly for the $\pi^0 \pi^0$ final state, we have
\bea
A_{00}&=&-\sqrt{ \frac{1}{3} } a_0 e^{i\, \delta_0} +
\sqrt{ \frac{2}{3} } a_2 e^{i\, \delta_2} \nonumber \\
\overline{A}{}_{00}&=&-\sqrt{ \frac{1}{3} } a_0^* e^{i\,\left( \delta_0 + \theta_{\rm
CP}-\overline{\theta}{}_{\rm T}\right)} +
\sqrt{ \frac{2}{3}} a_2^* e^{i\,\left( \delta_2 + \theta_{\rm
CP}-\overline{\theta}{}_{\rm T}\right)}~.
\nonumber\eea

As seen from the amplitudes, $B(\ks \rightarrow \pi^0
\pi^0)/B(\ks \rightarrow
\pi^+\pi^-)$ would be $0.5$ if $a_2=0$. Since the measured ratio is
$\sim 0.46$~\cite{Apostolakis:1999zw}, we can conclude that
$|a_2/a_0|<<1$.  It follows that
\ben
\frac{\overline{A}{}_{+-}}{A_{+-}} = \left(1 - 2 \epsilon ' \right)
e^{-i\, \left( 2\varphi_0+\overline{\theta}{}_{\rm
T}-\theta_{\rm CP} \right)}
\label{ampratio}\een
where the parameter $\epsilon '$ is given by
\ben
\epsilon ' = \frac{1}{\sqrt{2}} \left| \frac{a_2}{a_0}\right|
\sin(\varphi_2-\varphi_0) e^{i (\pi/2 + \delta_2-\delta_0)}
\label{epsilonp}\een
and $\varphi_{0,~2}=\arg a_{0,~2}$. 

As seen from equation \ref{ampratio}, CP violation in the decay amplitude,
$|A_{+-}|\neq |\overline{A}{}_{+-}|$, is present if $\Re \epsilon' \neq
0$. From equation~\ref{epsilonp}, this is possible only if 
\be
\sin(\varphi_2-\varphi_0)\neq 0~{\rm and}~\sin(\delta_2-\delta_0)\neq 0~.
\ee
i.e. both the weak and strong phases have to be different for the $I=0$
and $I=2$ decay amplitudes. More generally, there must be two processes
leading to the identical final state and both the strong and the weak
phases must be different between the two processes in order to generate CP
violation in the decay amplitudes. It should be noted that from the
measured $\pi$-$\pi$ scattering phase shift values, we
have~\cite{nakada-pipiphase}
\be
\arg \epsilon '= (43\pm6)^\circ
\ee

Using equations \ref{zetak} and \ref{ampratio}, it follows that
\bea
\zeta\frac{\overline{A}{}_{+-}}{A_{+-}}
&=&(1-2 \epsilon - 2\epsilon ' )e^{-i\, (\varphi_{\it
\Gamma}+2\varphi_0+\overline{\theta}{}_{\rm T}-\theta_{\rm CP})}
\nonumber \\
&\approx& 1-2 (\epsilon +\epsilon ')-i\,(\varphi_{\it
\Gamma}+2\varphi_0+\overline{\theta}{}_{\rm T}-\theta_{\rm CP})
\nonumber\eea
where the approximation is made assuming that the phase difference
between ${\it \Gamma}_{12}$ and $A_0\overline{A}_0$ is small, which will
be justified later. From equation \ref{etappdef}, $\eta_{+-}$ can be
derived to be
\be
\eta_{+-}=\epsilon + i(\varphi_{\it
\Gamma}+2\varphi_0+\overline{\theta}{}_{\rm T}-\theta_{\rm CP} )+ \epsilon'~.
\ee
Similarly the CP violation parameter for the $\pi^0 \pi^0$
decay channel, $\eta_{00}$, is given by
\be
\eta_{00}=\epsilon + i(\varphi_{\it
\Gamma}+2\varphi_0+\overline{\theta}{}_{\rm T}-\theta_{\rm CP} )- 2\epsilon '~.
\ee
Thus, we expect CP violation parameters to be different between
the $\pi^+ \pi^-$ and $\pi^0 \pi^0$ decay modes if $\epsilon ' \neq 0$.
It has been shown by four recent experiments, NA31~\cite{Barr:1993rx},
E731~\cite{Gibbons:1993zq}, KTeV~\cite{Alavi-Harati:1999xp} and
NA48~\cite{Fanti:1999nm}, 
\be
\left |\frac{\eta_{+-}}{\eta_{00}}\right|^2 = 1.0127\pm0.0028
\ee
i.e. CP violation in the decay amplitude is present in the neutral kaon
system. If we neglect $(\varphi_{\it
\Gamma}+2\varphi_0+\overline{\theta}{}_{\rm T}-\theta_{\rm CP} )$, it
follows that
\be
\Re \left( \frac{\epsilon '}{\epsilon} \right) = \frac{1}{6} \left( \left
|\frac{\eta_{+-}}{\eta_{00}}\right|^2 -1 \right)~.
\ee

\subsection{Phase of Decay Matrix}\label{sec-gamma}
As seen from equation~\ref{g-element}, evaluation of ${\it \Gamma}_{12}$
involves the decay final states which are common to $\ko$ and $\kob$,
which are $2\pi(I=0)$, $2\pi(I=2)$, $3\pi(I=1)$, $3\pi(I=2)$ and
$3\pi(I=3)$ states:
\be
{\it \Gamma}_{12} \approx \sum_{I=0,2}
A_{2\pi(I)}^* \overline{A}{}_{2\pi(I)}+\sum_{I=1,2,3}A_{3\pi(I)}^*
\overline{A}{}_{3\pi(I)}~.
\ee

The contribution
from the decay amplitude to the $2\pi(I=2)$ state is suppressed by the
$\Delta I =1/2$ rule and the small measured value of $\epsilon '$. The
contribution from the three-pion decay amplitudes are suppressed by
${\it
\Gamma}_{\rm L}/{\it
\Gamma}_{\rm S}$ and the measured upper limits for the CP violation
parameter for the $\pi^+ \pi^- \pi^0$ and $\pi^0 \pi^0 \pi^0$ final
states. In conclusion, the phase of
${\it
\Gamma}_{12}$ is essentially given by the phase of the $A_0$ amplitude,
and it can be expressed as 
\be
\varphi_{\it \Gamma}\approx \arg{A_0^* \overline{A}{}_0}
=-2\varphi_0-\overline{\theta}{}_{\rm T}+\theta_{\rm CP} 
\ee
so that
\be
|\varphi_{\it
\Gamma}+2\varphi_0+\overline{\theta}{}_{\rm T}-\theta_{\rm CP}| <
O(10^{-5})~.
\ee
Thus $|\varphi_{\it
\Gamma}+2\varphi_0+\overline{\theta}{}_{\rm T}-\theta_{\rm
CP}|<<|\epsilon|$, justifying the approximations made before.

\subsection{The Standard Model Description}
\begin{figure}
\begin{center}
\epsfig{file=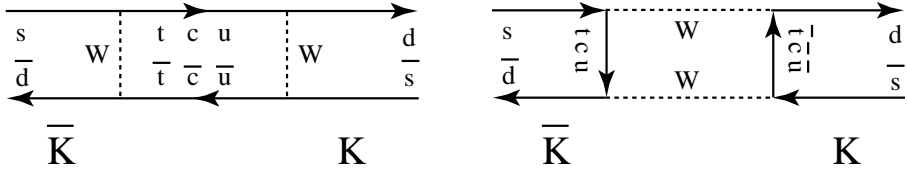,width=120mm}
\end{center}
\caption{The box diagrams contributing to the $\ko$-$\kob$ oscillations.}
\label{fig-box}
\end{figure}

\newcommand{\Vud}{V_{\rm ud}}
\newcommand{\Vus}{V_{\rm us}}
\newcommand{\Vub}{V_{\rm ub}}
\newcommand{\Vcd}{V_{\rm cd}}
\newcommand{\Vcs}{V_{\rm cs}}
\newcommand{\Vcb}{V_{\rm cb}}
\newcommand{\Vtd}{V_{\rm td}}
\newcommand{\Vts}{V_{\rm ts}}
\newcommand{\Vtb}{V_{\rm tb}}
In the framework of the Standard Model~\cite{nakada-buras}, the short
range contribution to $\ko$-$\kob$ oscillation is obtained from the box
diagrams (Figure~\ref{fig-box}) to be
\be
M_{12}^{\rm box}=- \frac{G_{\rm F}^2 f_{\rm K}^2 B_{\rm K} m_{\rm K}
m_{\rm W}^2 }{12\pi^2}\,\, \left[ \eta_1 \sigma_{\rm c}^2 S(x_{\rm c}) +
2
\eta_2
\sigma_{\rm c}
\sigma_{\rm t} E(x_{\rm c},x_{\rm t}) + \eta_3 \sigma_{\rm t}^2 S(x_{\rm
t})
\right] 
\ee
where $G_{\rm F}$ is the Fermi constant, $f_{\rm K}$, $B_{\rm K}$ and
$m_{\rm K}$ are the decay constant, $B$ parameter and mass
for the K-meson respectively and $m_{\rm W}$ is the mass of the
W-boson. The QCD correction factors are denoted by 
$\eta_1=1.38\pm0.20$, $\eta_2=0.57\pm0.01$ and $\eta_3=0.47\pm0.04$ and
$S$ and $E$ are known functions of the mass ratios, $x_{\rm
i}=m_{\rm i}^2/m_{\rm W}^2$ for top (i=t) and charm (i=c). Note that
\ben
S(x_{\rm c})\approx 2.4\times 10^{-4},~S(x_{\rm t})\approx 2.6,~
E(x_{\rm c},~x_{\rm c})\approx 2.2 \times 10^{-3}
\label{masscont}\een
for $m_{\rm c}=1.25~{\rm GeV}/c$, $m_{\rm t}=174~{\rm GeV}/c^2$ and $m_{\rm
W}=80~{\rm GeV}/c$~\cite{nakada-pdg}. The parameters
$\sigma_{\rm c}$ and
$\sigma_{\rm t}$ are the  combination of the elements of the
Cabibbo-Kobayashi-Maskawa quark mixing matrix (CKM-matrix),
\be
V_{\rm CKM}=\left( \begin{array}{ccc} \Vud & \Vus & \Vub \\
\Vcd & \Vcs & \Vcb \\
\Vtd & \Vts & \Vtb \end{array} \right)
\ee 
$\sigma_{\rm c}=V_{\rm cs} V_{\rm
cd}{}^*$ and
$\sigma_{\rm t}=V_{\rm ts} V_{\rm td}{}^*$. We adopt the following
approximation of the CKM matrix using the parameters introduced by
Wolfenstein~\cite{nakada-wolpara}:
\bea
V_{\rm CKM} \approx \left( \begin{array}{ccc}
1-\lambda^2/2 & \lambda & A \lambda^3 \left(\rho- i\, \eta \right) \\
-\lambda - i\, A^2 \lambda^5 \eta & 1-\lambda^2/2 & A \lambda^2 \\
A \lambda^3 \left( 1-\tilde{\rho} - i\,\tilde{\eta} \right) &
 -A \lambda^2- i \, A \lambda^4 \eta  & 1
\end{array} \right)
\label{ckmappr}\eea
where where $\tilde{\rho}=\rho(1-\lambda^2/2)$ and
$\tilde{\eta}=\eta(1-\lambda^2/2)$. The parameter $\lambda$ is known from
the light hadron decays to be
$0.221\pm0.002$. From the B-meson decays, $|\Vcb|=0.0402\pm0.0019$ and
$| \Vub/\Vcb|=0.090\pm0.025$ are measured~\cite{nakada-pdg}, giving
$A=0.823\pm0.042$ and
$\sqrt{\rho^2+\eta^2}=0.41\pm0.11$. The
$B$-parameter takes in account the difference between 
$
\langle 0 | H_{\rm W}  {\rm K^\pm} \rangle
$
and
$
\langle f | H_{\rm W} \kos
$
where $\langle 0 |$ is the hadronic vacuum state and $\langle f |$ is the
common quark states between $\ko$ and $\kob$. The theoretical evaluations
for this value vary between 0.5 and 1. 

In addition to $M_{12}^{\rm box}$, there are large
contributions from long range interactions
$M_{12}^{\rm LR}$, which are difficult to evaluate. Therefore, theoretical
predication for
$M_{12}=M_{12}^{\rm box}+M_{12}^{\rm LR}$ cannot be given.  The long
range interaction involves only the light flavours and its contribution
to $M_{12}$ is real in the CKM phase convention; the imaginary part of
$M_{12}$ is generated only by the box diagram. Therefore we can derive
\be
\sin(\varphi_M)=\frac{\Im M_{12}}{|M_{12}|} = \frac{\, 2\, \Im M_{12}^{\rm
box}\,}{\Delta m}~.
\ee
In the CKM phase convention, ${\it \Gamma}_{12}$ can be approximated as
real. Therefore, it follows that
\be
\Re \epsilon = -\frac{\,  \Im M_{12}^{\rm
box}\,}{2\, \Delta m}~.
\ee
Although there are considerable uncertainties to evaluate numerically
this expression, the currently allowed range of the Wolfenstein
parameters,
$\lambda$, $A$, $\rho$ and $\eta$ gives a consistent value of $\Re
\epsilon$ with the experimentally measured value.

\begin{figure}
\begin{center}
\epsfig{file=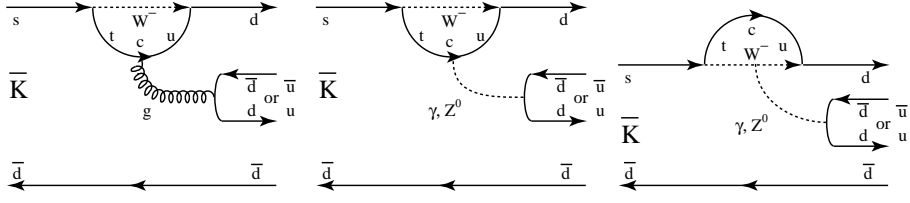,width=120mm}
\end{center}
\caption{Gluonic and electromagnetic penguins contributing to the $\kob
\rightarrow 2\pi$ decays.}
\label{fig-penguink}
\end{figure}

Prediction of $\epsilon '$ requires an accurate evaluation of the phase
difference between 
$a_{0}$ and $a_{2}$. For the $a_0$ amplitudes, the tree,
the gluonic penguin and the electroweak penguin diagrams contribute. Only
the tree and electroweak penguin diagrams make contributions to the
$a_2$ decay amplitude. All the penguin diagrams are shown in
Figure~\ref{fig-penguink}. Not only the short range interactions, but
also the hadronic matrix elements with long range interactions have to
be evaluated in the calculations. This makes the numerical determination
of
$\epsilon '$ very difficult. Within the theoretical
uncertainties, values of $\epsilon '$ calculated with the currently
allowed range of $\lambda$, $A$, $\rho$ and $\eta$ are consistent with the
data.

\subsection{CP Violation in Rare Decays}
\begin{figure}[b]
\begin{center}
\epsfig{file=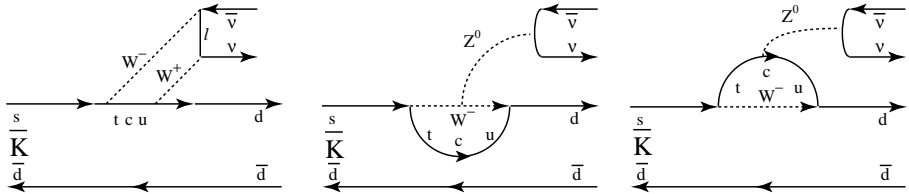,width=120mm}
\end{center}
\caption{The box and penguin diagrams generating $\kob \rightarrow \pi^0
\nu \overline{\nu}$ decays.}
\label{fig-knn}
\end{figure}

Experimental detection of $\kl \rightarrow \pi^0 \nu
\overline{\nu}$ is clearly very challenging. The final state is a CP
eigenstate with $CP=+1$. Therefore, observation of this decay is a sign
of CP violation. In the Standard Model, the decay is generated by penguin
diagrams or box diagrams as shown in Figure~\ref{fig-knn}. 

Since the final state
consists with only one hadron, long range strong interactions do not play
a role and the decay amplitudes can be denoted as
\bea
\langle \pi^0 \nu \overline{\nu} | H_{\rm W}  \kos &=& a_{\pi^0 \nu
\overline{\nu}}
\nonumber \\
\langle \pi^0 \nu \overline{\nu} | H_{\rm W}  \kobs &=& 
a_{\pi^0 \nu \overline{\nu}}^* \, e^{i\,\left( \theta_{\rm
CP}-\overline{\theta}{}_{\rm T}\right)}~.
\nonumber \eea 
Unlike for the $\ko \rightarrow 2\pi$ decays, $\phi_{\pi^0 \nu
\overline{\nu}}= \arg a_{\pi^0 \nu \overline{\nu}}$ could
be very different from $\phi_0$, so that we could have a situation
\bea
\left| \sin \left(\phi_{\it
\Gamma} + 2\, \phi_{\pi^0 \nu
\overline{\nu}}+\theta_{\rm
CP}-\overline{\theta}{}_{\rm T}\right) \right| &=&
\left| \sin \left(2\, \phi_{\pi^0 \nu
\overline{\nu}}- 2\phi_0 \right) \right| 
\nonumber \\
&>>& |\epsilon |~.
\nonumber \eea 
The $\kl$ decay
amplitude then becomes
\bea
\langle \pi^0 \nu \overline{\nu} | H_{\rm W} |\kl \rangle
&=& \frac{\, a_{\pi^0 \nu \overline{\nu}} \,}{\sqrt{\, 2 \,}} \left[ 1 -
(1- 2
\epsilon) e^{-i \, (2\, \phi_{\pi^0 \nu
\overline{\nu}}- 2\phi_0 )}\right]
\nonumber \\
&\approx& \sqrt{\, 2 \,} \, i\, |a_{\pi^0 \nu
\overline{\nu}}|\sin(2\, \phi_{\pi^0 \nu
\overline{\nu}}- 2\phi_0 )~. 
\nonumber \eea

Using isospin symmetry, the hadronic
matrix element of the $\ko \rightarrow \pi^0 \nu
\overline{\nu}$ decay amplitude and that of the $\rm K^+ \rightarrow \pi^+
e^+
\nu$ decay amplitudes can be related as 
\be
\langle \pi^0 | H_{\rm W} \kos = \langle \pi^0| H_{\rm W} | {\rm K}^+
\rangle~.
\ee
This allow us to express the branching fractions for $\kl \rightarrow
\pi^0 \nu \overline{\nu}$ using the branching fractions for $\rm K^+
\rightarrow \pi^0 e^+ \nu$ as~\cite{nakada-buras}
\bea
B(\kl \rightarrow
\pi^0 \nu \overline{\nu}) &=&\frac{\, |\langle \pi^0 \nu \overline{\nu} |
H_{\rm W} |\kl \rangle|^2 \,}{\Gamma_{\rm L}} \nonumber \\
&=&B({\rm K^+ \rightarrow \pi^0 e^+ \nu}) \frac{\tau_{\rm L}}{\tau_+}
\frac{3\, \alpha^2 \left[\Im (\Vts^* \Vtd) X(m_t) \right]^2}
{|\Vus|^2 2 \pi^2 \sin^4 \Theta_{\rm W}} \nonumber\\
&=&B({\rm K^+ \rightarrow \pi^0 e^+ \nu}) \frac{\tau_{\rm L}}{\tau_+}
\frac{3\, \alpha^2 \left[X(m_t) \right]^2}
{ 2 \pi^2 \sin^4 \Theta_{\rm W}}A^4 \lambda^8 (1-\lambda^2/2)^2 \eta^2
\nonumber \\
&\approx& 3\times 10^{-11}
\nonumber \eea
where $X$ is a known function and $\Theta_{\rm W}$ is the weak mixing
angle. Since the hadronic matrix element is taken from the data, the
theoretical uncertainties in this determination is very small. Also the
imaginary part of the amplitude is dominated by the short range 
interactions which can be reliably calculated. Therefore, the theoretical
prediction can be considered to be clean.

It is interesting to note that the CP violation parameter
\be
\eta_{\pi^0 \nu
\overline{\nu}} = \frac{\langle \pi^0 \nu
\overline{\nu}| V | \kl \rangle }{\langle \pi^0 \nu
\overline{\nu}| V | \ks \rangle}
\ee
as defined in the $2\pi$ case has $|\eta_{\pi^0 \nu
\overline{\nu}}|>>|\epsilon|$, although the both final states have
$CP=+1$.

The current experimental measurement for this branching fraction is
$< 5.9
\times 10^{-7}$ with $90\%$ confidence by the KTeV
experiment~\cite{Alavi-Harati:1999hd}, which is still far from the
expected number. However, there are several proposals to observe the
decays in the near future.

\section{B-meson System} 
\def\epm{\rm e^+ e^-}
\def\lumi{\rm cm^{-2} s^{-1}}
\newcommand{\jpsiks}{{\rm J/\psi \, K_S}}
\newcommand{\jpsiphi}{{\rm J/\psi \, \phi}}
\newcommand{\dstI}{{\rm D^{*-} \pi^+}}
\newcommand{\dstII}{{\rm D^{*+} \pi^-}}
\newcommand{\dmd}{{\it \Delta}m_{\rm d}}
\newcommand{\dms}{{\it \Delta}m_{\rm s}}
\newcommand{\dmK}{{\it \Delta}m_{\rm K}}
\newcommand{\bo}{{\rm B}^0}
\newcommand{\bob}{\overline{\rm B}{}^0}
\newcommand{\bso}{\rm B_s^0}
\newcommand{\bsob}{\overline{\rm B}{}_{\rm s}^0}
\newcommand{\bd}{{\rm B_d}}
\newcommand{\bs}{{\rm B_s}}
\newcommand{\bos}{|{\rm B}^0\rangle}
\newcommand{\bobs}{|\overline{\rm B}{}^0\rangle}
\newcommand{\bsos}{|\rm B_s^0\rangle}
\newcommand{\bsobs}{|\overline{\rm B}{}_{\rm s}^0\rangle}
\newcommand{\bl}{{\rm B_l}}
\newcommand{\bh}{{\rm B_h}}
\newcommand{\bls}{|{\rm B_l}\rangle}
\newcommand{\bhs}{|{\rm B_h}\rangle}
\subsection{The Standard Model Description}
\subsubsection{Some Elements of The CKM Matrix}
\begin{figure}
\begin{center}
\epsfig{file=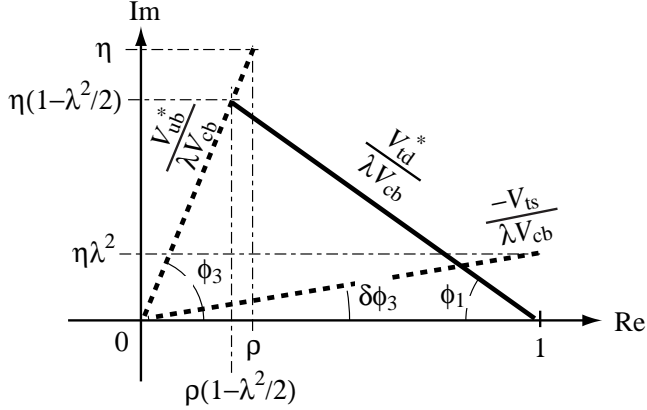,width=85mm}
\end{center}
\caption{Three elements of the CKM matrix, $\Vtd$, $\Vub$, and $\Vts$ and
the definitions of $\phi_1$, $\phi_3$ and $\delta \phi_3$. }
\label{fig-trangle}
\end{figure}

Among the nine elements of the CKM matrix, five of them related to the
third generation play important roles in the B meson system: $\Vtd$,
$\Vub$, $\Vts$, $\Vcb$ and $\Vtb$. In the approximation given in equation
\ref{ckmappr}, the  phases of the five elements are given by
\be
\arg \Vtd = - \phi_1,~\arg \Vub = - \phi_3,~\arg \Vts =  \delta \phi_3 +
\pi,~
\arg \Vcb =\arg \Vtb =0
\ee
where
\be
\phi_1 = \tan^{-1} \frac{\eta}{1-\rho}, \phi_3=\tan^{-1}
\frac{\eta}{\rho}, 
\delta \phi_3 = \tan^{-1} \lambda^2 \eta.
\ee
Figure~\ref{fig-trangle} shows the angles in $\rho$ and $\eta$
planes. Note that $\phi_1$ and $\phi_3$ are often referred to as 
$\beta$ and $\gamma$. Clearly
$\delta
\phi_3$ is very small,
$\sim 0.02$.

\subsubsection{Oscillation Amplitude}
In the Standard Model, $\rm B$-$\rm \overline{B}$ oscillation
is totally governed by the short range interactions, i.e. the box
diagrams. Furthermore, only the top quark plays a role in the box diagram
due to the large top quark mass (see equation \ref{masscont}) and the
structure of the CKM matrix;
\be
\frac{\Re(\Vtd^*\Vtb)}{\Re(\Vcd^*\Vcb)}=(\tilde{\rho}-1) \approx1, 
~~\frac{\Im(\Vtd^*\Vtb)}{\Im(\Vcd^*\Vcb)}\approx\frac{1}{\lambda^2}>>1
\ee
as seen from equation \ref{ckmappr}.

Therefore, the off diagonal element of the
mass matrix, $M_{12}$ is given by~\cite{nakada-buras}
\ben
M_{12}=-
\frac{G_{\rm F}^2 f_{\rm B_d}^2 B_{\rm B_d} m_{\rm B_d} m_{\rm W}^2
}{12
\pi^2} 
\eta_{\rm B_d} S(x_{\rm t}) (\Vtd^*\Vtb)^2 ~~~\mbox{for $\bd$}
\label{bdoscil}\een
where $f_{\rm B_d}$, $B_{\rm B_d}$ and $m_{\rm B_d}$ are the decay
constant, B-parameter and the mass of the $\rm B_d$ meson. 

Similarly for the $\rm B_s$ meson, we obtain
\be
M_{12}=-
\frac{G_{\rm F}^2 f_{\rm B_s}^2 B_{\rm B_s} m_{\rm B_s} m_{\rm W}^2
}{12
\pi^2} 
\eta_{\rm B_s} S(x_{\rm t}) (\Vts^*\Vtb)^2 ~~~\mbox{for $\bs$}
\ee
where $f_{\rm B_s}$, $B_{\rm B_s}$ and $m_{\rm B_s}$ are the decay
constant, B-parameter and the mass of the $\rm B_s$ meson.

The phase of $M_{12}$ is then given by
\be
\arg M_{12} = \left\{
\begin{array}{ll}
\arg (\Vtd^*\Vtb)^2 +\pi = 2\phi_1 + \pi& \mbox{for $\bd$} \\
 \arg (\Vts^*\Vtb)^2 +\pi = -2\delta \phi_3 + \pi &
\mbox{for $\bs$.}\rule{0mm}{5mm}
\end{array} \right.
\ee

The parameter ${\it \Gamma}_{12}$ can also be determined by taking the
absorptive part of the box diagrams with charm and up quarks in the
loops. The phase difference between $M_{12}$ and ${\it \Gamma}_{12}$ is
given by
\ben
\arg M_{12} - \arg {\it \Gamma}_{12} = \pi + 
\frac{\, 8\, }{3}\left( \frac{ m_{\rm c}}{m_{\rm b}} \right)^2 \eta
\times
\left\{
\begin{array}{cl}
 \frac{\rule[-1.5mm]{0mm}{4mm}\textstyle
1}{\rule{0mm}{4mm}\textstyle (1-\rho)^2 +
\eta^2} &:\bd
\vspace{ 2mm} \\
 \lambda^2  &:\bs \end{array} \right.
\label{phasemg}\een
i.e. $\sin (\arg M_{12} - \arg {\it \Gamma}_{12})$ is small for $\rm B_d$
and very small for $\rm B_s$. Note that $M_{12}$ and
$\itagamma_{12}$ are antiparallel. Therefore, the
approximations for
$\zeta$, $m_\pm$ and $\itagamma_\pm$ given on page~\pageref{page-approx}
are valid with $n=1$. Since we will rely on the Standard Model
description of
$M_{12}$, and our experimental knowledge of the decay amplitudes is still
limited, we adopt b)~$\phi_M$ base. We refer the mass eigenstate with
larger mass as $\bh$ (B-heavy) and the other $\bl$ (B-light) with their
masses and decay width are given by:
\be
m_{\rm h}= M+|M_{12}|,~ \itagamma_{\rm h}=\itagamma-|\itagamma_{12}|
\ee
and
\be
m_{\rm l}= M-|M_{12}|,~ \itagamma_{\rm l}=\itagamma+|\itagamma_{12}|
\ee
respectively, and $\bh$ ($\bl$) corresponds to $\rm P_+$ ($\rm P_-$)
defined in equation \ref{klks}.

For both $\bd$ and
$\bs$, we can now derive
\ben
\frac{\Delta {\it \Gamma}}{\Delta m} =\left|
\frac{ {\it
\Gamma}_{12}} {M_{12}}\right| \approx \frac{3 \pi m_{\rm b}^2}{2 m_{\rm
W}^2 S(x_{\rm t} )}\approx 5\times 10^{-3}\hspace{1cm} \mbox{for $\bd$
and $\bs$}
\label{moverg}\een
for $m_{\rm b}=4.25$ GeV, $m_{\rm W}=80$ GeV and $m_{\rm t}=174$ GeV,
where 
$\Delta m$ and $\Delta \itagamma$ are defined as positive: 
\be
\Delta m = m_{\rm h}-m_{\rm l}, \Delta \itagamma =  \itagamma_{\rm l} - 
\itagamma_{\rm h}~.
\ee
Using the measured values
of
$\Delta m = (0.464
\pm 0.018)
\times 10^{12}~\hbar {\rm s^{-1}}$ and the average lifetime
$\tau=1/\hat{\it \Gamma} =(1.54 \pm 0.03) \times 10^{-12}$~s for the $\bd$
mesons, where
$\hat{\it
\Gamma}$ is the averaged decay width, it follows that 
\be
\frac{\Delta {\it \Gamma}}{\hat{\it \Gamma}} \approx 4 \times
10^{-3}
\hspace{1cm} \mbox{for $\bd$}
\ee
and $\Delta {\it \Gamma}$ can be neglected in the decay time
distribution for the $\bd$ system. For the $\bs$ mesons, using the
measured lifetime $(1.54\pm0.07)\times 10^{-12}$~s, it follows that
\be
\frac{\Delta {\it \Gamma}}{\hat{\it \Gamma}} \approx 0.1
\hspace{1cm} \mbox{for $\bs$}.
\ee
The effect of $\Delta {\it \Gamma}$ is still not large, but can no
longer be neglected in the decay time distributions. 

The small decay width differences of the $\bd$ and $\bs$ systems do not
allow to separate one mass-eigenstate from the other, which can be done
for the kaon system by creating a
$\kl$ beam. Therefore, CP violation cannot be established by just
observing the decays as in the case of $\kl \rightarrow 2\pi$. We either
have to compare the decay rates of the initial $\bo$ and initial $\bob$
states or measure the time dependent decay rates of at least one of the
two cases, i.e. either initial $\bo$ or $\bob$. 

Since $\Delta m=2|M_{12}|$, one can extract 
\be
|\Vtd|^2=A^4 \lambda^6 \left[(1-\tilde{\rho})^2 +\tilde{\eta}^2 \right]
\ee
i.e. $\rho$ and $\eta$, from the measured $\bo$-$\bob$
oscillation frequency
$\dmd$ using equation~\ref{bdoscil}. However, theoretical uncertainties
in calculating the decay constant and B-parameter are considerable and
limit the accuracy on the extracted value of $|\Vtd|^2$. If the
$\bso$-$\bsob$ oscillation frequency $\dms=2|M_{12}^{\rm s}|$ is
measured, 
$|\Vtd|^2$ can be determined with much small uncertainty by
using the ratio $\dmd/\dms$, 
due to better controlled theoretical errors in $f_{\bd}/f_{\bs}$ and
$B_{\bd}/B_{\bs}$. However, the frequency of the
$\bso$-$\bsob$ oscillation is expected to be $> 1/\lambda^2 = 20$
times larger than that of the $\bo$-$\bob$ oscillation and we still
have to wait for sometime before it is measured.

Since $|M_{12}/{\it \Gamma}_{12}|<<1$, $\zeta$ given by
equation~\ref{mbase} can be further approximated as
\bea
\zeta \approx \left[1- \frac{1}{\, 2\,}\,  \Im  \left( \frac{\, {\it
\Gamma}_{12}\, }{ M_{12}} \right) \right] e^{ - i\,\varphi_M} 
\label{bzeta}\eea
where $\varphi_M = \arg M_{12}$ as before. 
Seen from equation \ref{phasemg} and \ref{moverg}, the
approximation $|\zeta|\approx 1$ is accurate to $10^{-3}$ or better. 

Similar to the kaon system, CP violation (and T violation) in the
oscillation can be measured from the time-dependent rate asymmetry
between the initial $\bob$ decaying into semileptonic final states
with $\rm e^+$ or $\mu^+$, $\overline{R}{}_+(t)$ and the initial
$\bo$ decaying into semileptonic final states with $\rm e^-$ or $\mu^-$,
$R_-(t)$. The asymmetry is given by
\be
\frac{\overline{R}{}_+(t)-R_-(t)}{\overline{R}{}_+(t)+R_-(t)}
= \frac{1-|\zeta|^4}{1+|\zeta|^4} 
\approx  O(10^{-3})~\mbox{for $\bd$ and } <<O(10^{-3})~\mbox{for $\bd$}
\ee
which is a very small signal.

From now on, we assume
\be
\zeta=e^{ - i\, \varphi_M}
\ee
for both $\bd$ and $\bs$ and $\Delta
{\it \Gamma}=0$ for $\bd$. 

In summary, the two mass
eigenstates are given by
\bea
\bhs &=& \frac{1}{\, \sqrt{2}\,} \left [|{\rm B}\rangle +
e^{ - i\,\varphi_M} |\overline{\rm B} \rangle \right]
\nonumber \\
\bls &=& \frac{1}{\, \sqrt{2}\,} \left [|{\rm B}\rangle -
e^{ - i\,\varphi_M} |\overline{\rm B} \rangle \right]
\nonumber \eea
and
\be
m_{\rm h}=m_0+|M_{12}|,~m_{\rm l}=m_0-|M_{12}|,~\Delta m =
m_{\rm h}-m_{\rm l}
\ee
for $\bd$ and $\bs$. For the decay width, we have
\be \begin{array}{ll} 
{\it \Gamma}_{\rm l}={\it \Gamma}_{\rm h}& \mbox{for $\bd$}
 \\
{\it \Gamma}_{\rm l}={\it \Gamma}_0+|{\it \Gamma}_{12}|,~ 
{\it \Gamma}_{\rm h}={\it \Gamma}_0-|{\it \Gamma}_{12}|,~\Delta {\it
\Gamma}={\it \Gamma}_{\rm l}-{\it \Gamma}_{\rm h} &
\mbox{for $\bs$}
\end{array} \ee

\subsubsection{Time Dependent Decay Rates}
Since $\Delta {\it \Gamma}$ is small in the B meson system, it is more
convenient to derive the time dependent decay rate from the
particle-antiparticle base rather than the mass eigenstate base. Using,
equations
\ref{eq-po1} and
\ref{eq-pob1} the time dependent decay rates for the final state f can be
derived as
\bea
R_{\rm f} (t) &\propto& \frac{\, | A_{\rm f} |^2 \,}{2}
e^{- \hat{\it \Gamma} \, t} \left[ I_+ (t) + I_- (t) \right]
\label{ptof}\\
\overline{R}_{\rm f} (t) &\propto& \frac{| A_{\rm f} |^2}{\, 2
|\zeta|^2
\,} e^{- \hat{\it \Gamma} \, t} \left[ I_+ (t) - I_- (t) \right]
\label{pbtof}
\eea
where $\hat{\it \Gamma}$ is the averaged decay time, $\hat{\it
\Gamma}=({\it \Gamma}_+ + {\it \Gamma}_-)/2$, and $A_{\rm f}$ is the
instantaneous decay amplitude for the $\po \rightarrow {\rm f}$ decays.
The two time dependent functions, $I_+(t)$ and $I_-(t)$, are given by
\bea
I_+(t)&=&(1+|L_{\rm f}|^2)\, \cosh \frac{\Delta {\it \Gamma} }{2} \, t
+ 2 \Re L_{\rm f} \, \sinh \frac{\Delta {\it \Gamma} }{2} \, t \nonumber
\\
I_-(t)&=&(1-|L_{\rm f}|^2) \cos \Delta m \, t
+ 2 \Im L_{\rm f} \, \sin \Delta m \, t ~.\nonumber
\eea
The parameter $L_{\rm f}$ is given by
\be
L_{\rm f} = \zeta \frac{\, \overline{A}_{\rm f} \,}{A_{\rm f}}
\ee
where $\overline{A}_{\rm f}$ is the instantaneous decay amplitude for 
the $\pob \rightarrow {\rm f}$ decays.

The time dependent decay rate for the CP conjugated final states $\rm f^{
CP}$ are derived to be
\bea
\overline{R}_{\rm f^{CP}} (t) &\propto& \frac{\, | \overline{A}_{\rm
f^{CP}} |^2
\,}{2} e^{- \overline{{\it \Gamma}} \, t} \left[ \overline{I}_+^{\rm CP}
(t) + \overline{I}_-^{\rm CP} (t)
\right]
\label{pbtofcp}\\
R_{\rm f^{CP}} (t) &\propto& \frac{\, | \overline{A}_{\rm f^{CP}} |^2
|\zeta|^2 \,}{2} e^{- \overline{{\it \Gamma}} \, t} \left[
\overline{I}_+^{\rm CP} (t) - \overline{I}_-^{\rm CP} (t)
\right]
\label{btofcp}
\eea
where $\overline{A}_{\rm f^{CP}}$ is the instantaneous decay amplitude for
the
$\pob
\rightarrow {\rm f^{CP}}$ decays. Two time dependent decay rates,
$\overline{I}_+^{\rm CP}(t)$ and $\overline{I}_-^{\rm CP}(t)$ are given by
\bea
\overline{I}^{\rm CP}_+(t)&=&(1+|L_{\rm f}^{\rm CP}|^2)\, \cosh
\frac{\Delta {\it
\Gamma} }{2}
\, t + 2 \Re L_{\rm f}^{\rm CP}  \sinh \frac{\Delta {\it \Gamma}
}{2}
\, t
\nonumber
\\
\overline{I}^{\rm CP}_-(t)&=&(1-|L_{\rm f}^{\rm CP}|^2) \cos \Delta
m
\, t + 2 \Im L_{\rm f}^{\rm CP}  \sin \Delta m \, t \nonumber
\eea
where the parameter, $L_{\rm f}^{\rm CP}$, is given by
\be
L_{\rm f}^{\rm CP} = \frac{1}{\, \zeta \,} \frac{A_{\rm
f^{CP}}}{\, \overline{A}_{\rm f^{CP}} \,}
\ee
and $A_{\rm f^{CP}}$ is the instantaneous decay amplitude for
the $\po \rightarrow {\rm f^{CP}}$ decays. 

The decay rates $R_f(t)$ and $\overline{R}_{\rm f^{CP}}(t)$ are CP
conjugate to each other and so are $\overline{R}_f(t)$ and $R_{\rm
f^{CP}}(t)$. If there exists any difference between the CP conjugated
processes, this is a clear sign of CP violation.

The final state f can be classified into the
following four different cases:
\begin{enumerate}
\item[I.] Flavour specific final state 
($A_{\rm f}=\overline{A}_{\rm f^{CP}}=0$ or $A_{\rm
f^{CP}}=\overline{A}_{\rm f}=0$)
\item[II.] Flavour non specific final state
\begin{enumerate}
\item[II-a.] CP eigenstate ($A_{\rm f}=A_{\rm f^{CP}}$ and 
$\overline{A}_{\rm f}=\overline{A}_{\rm f^{CP}}$)
\item[II-b.] mixed CP eigenstate ($A_{\rm f}=A_{\rm f^{CP}}$ and 
$\overline{A}_{\rm f}=\overline{A}_{\rm f^{CP}}$)
\item[II-c.] CP non eigenstate
\end{enumerate}
\end{enumerate}

\subsubsection{CP Violation: Clean Case}
\begin{figure}
\begin{center}
\epsfig{file=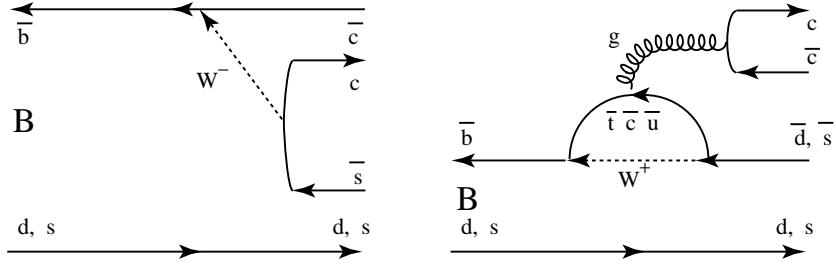,width=110mm}
\end{center}
\caption{Tree and penguin diagrams contributing to the $\bo \rightarrow
\jpsiks$ and
$\bso\rightarrow \jpsiphi$ decays.}
\label{fig-psiks}
\end{figure}

The contribution to the $\bo$ decaying into $\jpsiks$ is dominated by the
tree diagram with $\Vcb^* \Vcs$. Although there exist some contribution
from the penguin diagrams, the dominant penguin diagram contribution has
the CKM phase $\Vtb^* \Vts$ which is close to that of the tree diagram
(Figure~\ref{fig-psiks}). Thus, we can safely assume that there is no CP
violation in the decay amplitude and the ratio of the $\bob$ and $\bo$
decay amplitudes is given only by the CKM part. By noting that
$CP(\jpsiks)=-1$ we obtain
\be
\frac{A(\bob \rightarrow \jpsiks)}{A(\bo \rightarrow \jpsiks)}
=-\frac{ (\Vcb^* \Vcs \Vus^* \Vud)^2}{|\Vcb^* \Vcs \Vus^* \Vud|^2}~.
\ee
Using the formulae developed in the previous section, the time dependent
rates for the initial
$\bo$ decaying into
$\jpsiks$,
$R_{\jpsiks}(t)$, and that for $\bob$ decaying into $\jpsiks$,
$\overline{R}{}_{\jpsiks}(t)$ are given by
\bea
R_{\jpsiks}(t) \propto e^{- \hat{\it \Gamma} t} \left(
1 + \Im\,  L_{\jpsiks} \sin \Delta m \, t \right) \nonumber \\
\overline{R}{}_{\jpsiks}(t) \propto e^{- \hat{\it \Gamma} t} \left(
1 -  \Im\,  L_{\jpsiks} \sin \Delta m \, t \right) 
\nonumber \eea
which allow to extract
\be
\Im\,  L_{\jpsiks}= \Im\, \left( \zeta \times \frac{A(\bob \rightarrow
\jpsiks)}{A(\bo
\rightarrow
\jpsiks)} \right) =-  \Im\, \left[\frac{ (\Vtd^* \Vtb \Vcb^* \Vcs \Vus^*
\Vud)^2}{|\Vtd^* \Vtb \Vcb^*
\Vcs
\Vus^* \Vud|^2} \right]
\ee
With the Wolfenstein parameterization, it follows that
\be
\Im\,  L_{\jpsiks}= - \sin 2\phi_1~.
\ee 

The same argument holds for the $\bs \rightarrow \jpsiphi$ decays and
from the time dependent decay rates
\bea
R_{\jpsiphi}(t) \propto e^{- \hat {\it \Gamma} t} \left(
\cosh \frac{\Delta \Gamma}{2}t + 2 \Re\, L_{\jpsiphi} \sinh \frac{\Delta
\Gamma}{2}t + \Im\, L_{\jpsiphi} \sin \Delta m \, t \right) \nonumber \\
\overline{R}{}_{\jpsiphi}(t) \propto e^{- \hat {\it \Gamma} t} \left(
\cosh \frac{\Delta \Gamma}{2}t + 2 \Re\, L_{\jpsiphi} \sinh \frac{\Delta
\Gamma}{2}t - \Im, L_{\jpsiphi} \sin \Delta m \, t \right) 
\nonumber \eea
one can extract
\be
\Im \,  L_{\jpsiphi}= \Im \left[ \zeta \times \frac{A(\bsob \rightarrow
\jpsiphi)}{A(\bso
\rightarrow
\jpsiphi)}\right] = - \sin 2\delta \phi_3 
\ee

Note that we assumed in the calculation above that
$CP(\jpsiphi)=+1$, i.e. the
$\rm J/\psi\, \phi$ state is in the lowest orbital angular momentum state
of $l =0$. If there exists the $l=1$ state with $CP(\jpsiphi)=-1$,
the measured $\Im \,  L_{\jpsiphi}$ will be diluted and the fraction of
the $CP=-1$ state must be experimentally measured. If there
is the same amount of
$CP=+1$ state and
$CP=-1$ state,
$\Im
\, 
L_{\jpsiphi}$ will vanish.

\begin{figure}
\begin{center}
\epsfig{file=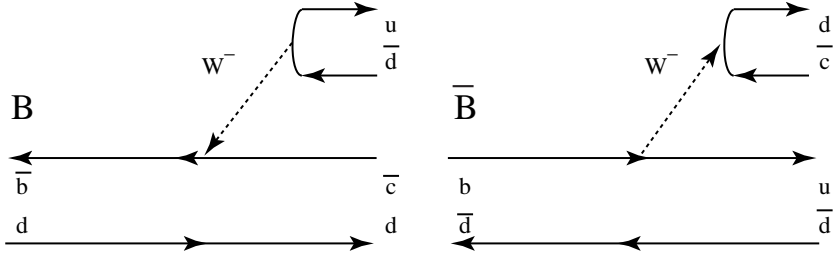,width=110mm}
\end{center}
\caption{Tree diagrams contributing for the $\bo \rightarrow \dstI$ and
$\bob \rightarrow \dstI$ decays.}
\label{fig-dsk}
\end{figure}

An even cleaner decay channel is $\bo \rightarrow {\rm D^{*\mp}}\pi^\pm$.
There is only one tree diagram, $\rm \overline{b} \rightarrow
\overline{c} + W^{+}$ followed by $\rm W^+ \rightarrow u+\overline{d}$,
which contributes to the
$\bo \rightarrow \dstI$ decays. The same final state can be produced from
the $\bob$ decays with another tree diagram, $\rm b \rightarrow
u + W^-$ followed by $\rm W^- \rightarrow \overline{c}+d$
(Figure~\ref{fig-dsk}).  Therefore, the time dependent rate for the
initial
$\bo$ decaying into
$\dstI$ is given by
\be
R_{\rm D^{*-}}(t) \propto
e^{- \hat {\it \Gamma} t} \left[ 1 + 
\frac{(1- |L_{\dstI}|^2)}{(1+|L_{\dstI}|^2)} \cos \Delta m\, t
+ \frac{2\Im L_{\dstI}}{(1+ |L_{\dstI}|^2)} \sin \Delta m\, t \right]
\ee
where
\be
L_{\dstI} = \zeta \times \frac{A(\bob \rightarrow \dstI)}{A(\bo
\rightarrow
\dstI)}
\ee
The weak phase of $A(\bob \rightarrow \dstI)$ is given by $\Vub\Vcd^*$
and that of $A(\bo \rightarrow
\dstI)$ by $\Vcb^* \Vud$. The phase of
$L_{\dstI}$ is then derived to be 
\bea
\arg L_{\dstI} &=& \arg \Vub - \arg M_{12} + \varphi_{\rm S}
\nonumber \\
&=& -\phi_3 + 2\phi_1 + \varphi_{\rm S}
\nonumber \eea
where $\varphi_{\rm S}$ is a possible strong phase difference between
the $\rm b\rightarrow u+W^-$ and $\rm \overline{b} \rightarrow
\overline{c}+W^+$ tree diagrams. 

CP conjugated decay amplitudes of $A(\bob \rightarrow \dstI)$ and $A(\bo
\rightarrow
\dstI)$, i.e. $A(\bo \rightarrow \dstII)$ and
$A(\bob
\rightarrow
\dstII)$ respectively, are obtained by taking the complex conjugate of the
weak amplitudes while the strong phase remains unchanged. Thus for
$\dstII$ we obtain
\be
R_{\rm D^{*+}}(t) \propto
e^{- \hat {\it \Gamma} t} \left[ 1 - 
\frac{(1- |L_{\dstI}^{\rm CP}|^2)}{(1+|L_{\dstI}^{\rm CP}|^2)} \cos
\Delta m\, t - \frac{2\Im L_{\dstI}^{\rm CP}}{(1+ |L_{\dstI}^{\rm
CP}|^2)} \sin
\Delta m\, t
\right]
\ee
where
\be
L_{\dstI}^{\rm CP} = \frac{1}{\zeta} \times \frac{A(\bo \rightarrow
\dstII)}{A(\bob
\rightarrow
\dstII)}
\ee
and the phase of $L_{\dstI}^{\rm CP}$ is given by
\bea
\arg L_{\dstI}^{\rm CP} &=& -\arg \Vub + \arg M_{12} + \varphi_{\rm S}
\nonumber \\
&=& \phi_3 - 2\phi_1 + \varphi_{\rm S}
\nonumber \eea
From the two time dependent decay rates, we can extract $\phi_3-2\phi_1$.

Note that 
\be
\left| L_{\dstI} \right | = \left| L_{\dstI}^{\rm CP} \right | \approx
\left| \frac{\Vub \Vcd^*}{\Vcb^* \Vud}\right|=\lambda^2\sqrt{\rho^2\eta^2}
<<1
\ee
i.e. the effect we have to measure is small.

The CP conjugated time
dependent decay rate distributions are given by
\be
\overline{R}_{\rm D^{*+}}(t) \propto
e^{- \hat {\it \Gamma} t} \left[ 1 + 
\frac{(1- |L_{\dstI}^{\rm CP}|^2)}{(1+|L_{\dstI}^{\rm CP}|^2)} \cos
\Delta m\, t + \frac{2\Im L_{\dstI}^{\rm CP}}{(1+ |L_{\dstI}^{\rm
CP}|^2)} \sin
\Delta m\, t
\right]
\ee
and
\be
\overline{R}_{\rm D^{*-}}(t) \propto
e^{- \hat {\it \Gamma} t} \left[ 1 - 
\frac{(1- |L_{\dstI}|^2)}{(1+|L_{\dstI}|^2)} \cos \Delta m\, t
- \frac{2\Im L_{\dstI}}{(1+ |L_{\dstI}|^2)} \sin \Delta m\, t \right]
\ee
which can be used to obtain the same information.

A similar method can be used for the $\bso \rightarrow {\rm D_s^\mp
K^\pm}$ decays to extract $\phi_3 - 2 \delta \phi_3$. The effect is
larger since
\be
|L_{\rm D_s^- K^+}| \approx \left| \frac{\Vub \Vcs^*}{\Vcb^* \Vus}
\right| = \sqrt{\rho^2 + \eta^2}= O(1)~.
\ee

\subsubsection{CP Violation: Not So Clean Case}
The penguin contribution to the $\bd \rightarrow \pi^+ \pi^-$ decay was
originally thought to be small and the decay would be dominated by the
$\rm b \rightarrow u+W$ tree diagram. However, the discovery of $B(\bd
\rightarrow {\rm K}^\pm \pi^\mp)>B(\bd \rightarrow \pi^+ \pi^-)$
indicates that the contribution of the penguin diagrams to the
$\bd \rightarrow \pi^+ \pi^-$ amplitude should be $\sim 20\%$ or more.

Due to the penguin contribution, the phase of the $\bo \rightarrow
\pi^+ \pi^-$ decay amplitude deviates from that of $\Vub^*$.
Furthermore, CP violation in the decay amplitude could be present.
Evaluation of those effects involves calculating
contributions from different diagrams accurately. Strong interactions may
play an important role as well. Therefore, this decay mode may not be
ideal to make precise determinations of $\rho$ and $\eta$ from CP
violation. 

\subsection{Case with New Physics}
Decay processes where only the tree diagrams contribute should be
unaffected by the presence of physics beyond the Standard
Model. Therefore, $|\Vcb|$ and $|\Vub|$ obtained from the semileptonic
decays of B mesons would not be affected by the new physics and $A$ and
$\rho^2 + \eta^2$ can be obtained even if physics beyond the Standard
Model is present.

New physics could generate $\bo$-$\bob$ and
$\bso$-$\bsob$ oscillations by new particles generating new box
diagrams. They could also generate a tree level flavour changing neutral
current contributing to the oscillation. Since these
contributions are through ``virtual'' states, they contribute to $M_{12}$
with little effect on ${\it
\Gamma}_{12}$, i.e. 
\be
M_{12}=M_{12}^{\rm SM}+M_{12}^{\rm NP},~{\it \Gamma}_{12}={\it
\Gamma}_{12}^{\rm SM}
\ee
where $M_{12}^{\rm SM}$ and ${\it
\Gamma}_{12}^{\rm SM}$ are due to the Standard Model and
$M_{12}^{\rm NP}$ is the contribution from the new physics. The
measured $\Delta m$ is given by $2|M_{12}|$ and can no longer used to
extract $|\Vtd|^2$ due to $M_{12}^{\rm NP}$.

Since
\be
\left| \frac{{\it \Gamma}_{12}}{M_{12}} \right|
= \frac{2\left|{\it \Gamma}_{12}^{\rm SM}\right|}{\Delta m} 
\ee
remains small, CP violation in the oscillation remains small
as seen from equation~\ref{bzeta}. Therefore, 
\be
\zeta=e^{ - i\,\varphi_M}
\ee 
is still valid. However, note that 
\be
\varphi_M \equiv\arg M_{12} \neq \arg M_{12}^{\rm SM}.
\ee

Decay amplitudes from the penguin diagrams can be affected by physics
beyond the Standard Model since new particles can contribute virtually in
the loop. Therefore, the modes such as $\bd$ decaying into $\pi^+ \pi^-$,
$\rm K^\pm \pi^\mp$ may have some contribution from the new physics.

Since the decays $\bd \rightarrow \jpsiks$ and $\bs
\rightarrow \jpsiphi$ are tree dominated, they are little affected by new
physics. Therefore we have
\be
\frac{A(\bob \rightarrow \jpsiks)}{A(\bo \rightarrow \jpsiks)}
=-\frac{A(\bso \rightarrow \jpsiphi)}{A(\bsob \rightarrow \jpsiphi)}
=-1
\ee
with the phase convention due to the Wolfenstein parameterization and 
\be
L_{\jpsiks,~ \jpsiphi} = \mp e^{ -i\varphi_M}~:
\mbox{$-$ for $\bd\rightarrow \jpsiks$ and $+$ for $\bs\rightarrow
\jpsiphi$}
\ee
and studies of the time dependent decay rates give $\arg
M_{12}$.

The $\rm B_d \rightarrow D^* \pi$ and $\rm B_s \rightarrow D_s K$ decays
are generated by only the tree diagrams and not affected by new physics.
Therefore we have
\be
\arg L_{\dstI} = -\phi_3 - \arg M_{12} + \varphi_{\rm S} 
\nonumber \ee
and
\be
\arg L_{\dstII} = \phi_3 + \arg M_{12} + \varphi_{\rm S}
\ee
and studies of the time dependent decay rates provide $\arg M_{12} +
\phi_3$. Similarly studies can be done for $\rm B_s \rightarrow D_s
K$.

By combining the measurements of $\bd \rightarrow \jpsiks$ and
$\rm \rightarrow D^* \pi$ or $\bs \rightarrow \jpsiphi$ and
$\rm \rightarrow D_s K$, the angle $\phi_3$ can be determined even with
presence of physics beyond the Standard Model. By comparing the result from
$\bd$ and that from $\bs$, consistency of the method can be tested. Since
the phase of $\Vub$ is given by
$\phi_3$ and its modulus is measured from the semileptonic decay, $\rho$
and $\eta$ can be extracted. Once $\lambda$, $A$, $\rho$ and $\eta$ are
known,
$M_{12}^{\rm SM}$ can be calculated and from the measured $\Delta m$ and
$\arg M_{12}$, the new physics contribution $M_{12}^{\rm NP}$ is
obtained. This can be used to identify the nature of the new physics
contributing to the oscillation.

\subsection{Experimental Prospects}
A possible experimental programme for the study of CP violation in the B
meson system and search for physics beyond the Standard Model can be
summarised in the following steps:
\begin{enumerate}
\item Determination of $|\Vcb|$ and $|\Vub|$ from semileptonic (and some
hadronic) decays.
\item Measurement of $\Delta m$ for $\bd$ and $\bs$
\item Measurement of $\Im L_{\jpsiks}$
\item Measurement of $L_{\jpsiphi}$, $L_{\rm D^{*\mp}\pi^\pm}$ and
$L_{\rm D_s^\mp K^\pm}$
\end{enumerate}

The first step has been made by ARGUS and CLEO at $\Upsilon(4S)$
machines and the four LEP experiments. BABAR and BELLE at the new
asymmetric
$\Upsilon(4S)$ machines and CLEO will improve the precisions on those
determinations. Future improvement of theory is also an important factor.
Half of the second step, $\Delta m(\bd)$ was done by ARGUS, CLEO, UA1
at the SPS Collider, the four LEP experiments, SLD at SLC and CDF at
the Tevatron. For
$\Delta m(\bs)$, we may have to wait for the next data taking by CDF, 
D0 and HERA-B. The third step will be made by BABAR, BELLE, CDF, D0 and
possibly HERA-B by the year 2005. 

After the second step, four parameters of the CKM matrix are all defined
within the framework of the Standard Model, e.g. $A$, $\lambda$, $\rho$
and $\eta$. The third step provides an additional information $\tan^{-1}
\eta/(1-\rho)$ within the framework of the Standard Model and 
consistency of the CKM picture can now be tested.

As demonstrated in the previous chapter, if physics beyond the Standard
Model exists, the fourth step is needed to clearly establish the
evidence of new physics and separate the effect due to the Standard Model
and that from new physics. After the third step, only $\rho^2+\eta^2$
will be known from $|\Vub|$ and the information on
$\tan^{-1}\eta/(1-\rho)$ is spoiled by new physics. Only after the fourth
step, $\rho$ and $\eta$ can be determined, together with isolating the new
physics contribution.

For the last step, new generation of experiments with statistics
much higher than $10^{10}$ B mesons are needed. The $\bs$ meson is an
essential ingredient. After 2005, LHC will be the most powerful source of
B mesons. Experiments must be equipped with a trigger efficient for
hadronic decay modes to gain high statistics for the necessary final
states. Particle identification is also crucial in order to reduce
background. LHCb is a detector at the LHC optimised for CP violation
studies with B mesons. The two general purpose LHC detectors, ATLAS and
CMS can contribute only to a limited aspect of the fourth step. A proposed
experiment at Tevatron, BTeV, can also make the last two steps.

Clearly CP violation is expected in many other decay channels. For many of
them, there are some theoretical problems for making accurate predictions.
However, they can be used to make a systematic study which will provide a
global picture whether CP violation can fit into the CKM picture. With
all those experiments, we continue to improve our understanding of CP
violation and hope to discover physics beyond the Standard Model.

\subsubsection*{Acknowledgement}
The author is very grateful to the organizers of this school for their
extended hospitality and efforts to prepare such a stimulating
environment. R.~Forty is acknowledged for reading this manuscript
and giving many useful comments. The author appreciates many useful comments
by O.~Schneider.

\end{document}